\documentclass[12pt]{article}

\usepackage{epsfig,multicol,multirow,cite}
\usepackage{amsmath,subfigure,latexsym,amssymb}

\newcommand{\be}{\begin{equation}}
\newcommand{\ee}{\end{equation}}
\newcommand{\nn}{\nonumber}
\newcommand{\bea}{\begin{eqnarray}}
\newcommand{\eea}{\end{eqnarray}} 

\newcommand{\la}{\langle}
\newcommand{\ra}{\rangle}

\newcommand{\Z}{\mathbb{Z}}
\newcommand{\R}{{\kern+.25em\sf{R}\kern-.78em\sf{I} \kern+.78em\kern-.25em}}
\newcommand{\RR}{{\kern+.25em\sf{R}\kern-.6em\sf{I} \kern+.6em\kern-.25em}}
\newcommand{\N}{{\kern+.25em\sf{N}\kern-.78em\sf{I} \kern+.78em\kern-.25em}}
\newcommand{\C}{{\kern+.25em\sf{C}\kern-.50em\sf{I} \kern+.50em\kern-.25em}}

\newcommand{\ri}{{\rm i}}

\newcommand{\vp}{\varphi}

\makeatletter
\@addtoreset{equation}{section}
\makeatother

\begin{document}
\begin{flushright}
CERN-PH-TH-2015-219
\end{flushright}
\vspace*{2mm}
 
\begin{center}
{\Large\bf Topological Susceptibility from Slabs}

\vspace*{7mm}

Wolfgang Bietenholz$^{\rm \, a}$, Philippe de Forcrand$^{\rm \, b,c}$
and Urs Gerber$^{\rm \, a,d}$\\ 
\ \\
$^{\rm \, a}$  Instituto de Ciencias Nucleares \\
Universidad Nacional Aut\'{o}noma de M\'{e}xico \\
A.P. 70-543, C.P. 04510 Distrito Federal, Mexico \\
\ \\
$^{\rm \, b}$
Institute for Theoretical Physics, ETH Z\"{u}rich\\
CH-8093 Z\"{u}rich, Switzerland\\
\ \\
$^{\rm \, c}$ CERN, Physics Department, TH Unit \\
CH-1211 Geneva 23, Switzerland\\
\ \\
$^{\rm \, d}$ Instituto de F\'{\i}sica y Matem\'{a}ticas\\
Universidad Michoacana de San Nicol\'{a}s de Hidalgo\\
Edificio C-3, Apdo. Postal 2-82 \\
C.P. 58040, Morelia, Michoac\'{a}n, Mexico
\end{center}

\vspace*{5mm}

\noindent
In quantum field theories with topological sectors, a 
non-perturbative quantity of interest is the topological 
susceptibility $\chi_{\rm t}$. In principle it seems straightforward
to measure $\chi_{\rm t}$ by means of Monte Carlo simulations. 
However, for local update algorithms and fine lattice spacings, 
this tends to be difficult, since the Monte Carlo history rarely 
changes the topological sector. 
Here we test a method to measure 
$\chi_{\rm t}$ even if data from only one sector are available.
It is based on the topological charges in sub-volumes, which 
we denote as slabs. Assuming a Gaussian distribution of these 
charges, this method enables the evaluation of $\chi_{\rm t}$, as we 
demonstrate with numerical results for non-linear $\sigma$-models.

\newpage

\section{The topological susceptibility $\chi_{\rm t}$}

There are a number of models in quantum field theory,
which have the property that the configurations occur 
in distinct topological sectors; each sector
is characterised by a topological charge $Q \in \Z$.
This refers either to infinite volume and configurations with
finite actions, or to finite volume with periodic boundary 
conditions. Here we consider the latter setting, with some 
volume $V$ in Euclidean space.

The models with this property include in particular 4d $SU(N)$ 
Yang-Mills theories for all $N \geq 2$. Fermions may be 
present as well, so this class of models encompasses QCD.
In that case, the quenched value of the topological susceptibility
$\chi_{\rm t}$ has a prominent application in the explanation of 
the heavy mass of the pseudo-scalar $\eta'$-meson \cite{WV}.

The measurement of $\chi_{\rm t}$ is a non-perturbative issue,
hence lattice simulations are the appropriate method. 
If the Monte Carlo history changes $Q$
frequently, this measurement is straightforward. However, for
most of the popular algorithms, including local update algorithms,
as well as the Hybrid Monte Carlo algorithm 
(which is standard in QCD with dynamical quarks),
the auto-correlation time with respect to $Q$ tends to be very
long, in particular on fine lattices.\footnote{In general, the
topological sectors are well-defined only in the continuum limit
(there are exceptions for lattice actions with a constraint that 
only admits very smooth configurations \cite{admis,topact,topactU1}). 
So in general transitions are only enabled by lattice artifacts, 
and by discrete jumps of the algorithm.}
This problem is getting even worse in the presence of chiral fermions. 

Recently, several indirect methods to measure $\chi_{\rm t}$ 
nevertheless have been suggested and tested \cite{othermethods,qqpap}.
Here we address a different approach for this purpose,
which was first sketched in Ref. \cite{Phil}, but which has
not been explored ever since (though a similar approach was
studied last year \cite{LSD}). It is described in Section 2.
Sections 3 and 4 give results for the 1d $O(2)$ and the 2d $O(3)$
model, respectively. Our conclusions are discussed in Section 5,
and an appendix is devoted to analytical considerations about 
artifacts in $\chi_{\rm t}$.

\section{Evaluating $\chi_{\rm t}$ from slabs}

We assume parity symmetry to hold, which implies 
$\la Q \ra = 0$. We further assume the topological
charges to obey a Gaussian probability 
distribution,\footnote{The assumption of a Gaussian distribution
of the topological charges appears natural in light of instanton
gas models. In case of the quantum rotor in infinite volume,
it can be demonstrated analytically \cite{rot97}. Simulations
confirm that it also holds --- at least to a good approximation ---
in the 2d $O(3)$ model \cite{qqpap} (cf.\ Section 4), 
and in QCD \cite{topGaus}.}
\be
p (Q) \propto \exp \left( - Q^{2} / (2 \chi_{\rm t} V) \right)
\ , \quad \chi_{\rm t} = \frac{1}{V} \, \la Q^{2} \ra \ .
\ee

The idea of the method that we are going to explore,
is to divide the periodic volume $V$ 
into sub-volumes, which we denote as {\em slabs},
and to extract $\chi_{\rm t}$ from the fluctuations of the topological 
charge within these slabs. This has the potential of providing
the result even based on configurations of a single 
topological sector (with respect to the entire volume $V$).

We consider just two slabs, of sizes $xV$ and $(1-x)V$, 
where $0 <x < 1$.\footnote{The extension of this method to a larger
number of slabs is straightforward, but hardly promising, since the
slab volumes should not be too small.} 
At fixed charge $Q$, we denote the topological 
charge contribution in the first slab as $q \in \R$; it is obtained
by integrating the topological charge density over the slab volume.
In general it is not an integer, because not all slab boundaries
are periodic. Thus we obtain the probability for the 
charge contributions $q$ and $Q-q$ in the two slabs,
\bea
p_{1}(q) \cdot p_{2}(Q-q) \vert_{x,V} & \propto &
\exp \Big( - \frac{q^{2}}{2 \chi_{\rm t}V x} \Big) \cdot
\exp \Big( - \frac{(Q-q)^{2}}{2 \chi_{\rm t}V (1-x)} \Big) \nn \\
& = & \exp \Big( - \frac{1}{2 \chi_{\rm t}V} \Big[ 
\frac{q{'}^{~2}}{x(1-x)} + Q^{2} \Big] \Big) \nn \\
 & \propto & \exp \Big( - \frac{1}{2 \chi_{\rm t}V}
\frac{q{'}^{\, 2}}{x(1-x)} \Big) \ ,
\label{p1p2}
\eea
where we defined $q' := q - xQ$. 

Eq.\ (\ref{p1p2}) implies $\la q \ra = xQ$, and therefore
\be
\la q{'}^{\, 2} \ra = \la q^{2} \ra - x^{2} Q^{2} \ .
\label{q2}
\ee
Thus, if we measure $\la q^{2} \ra$ at a set of $x$ values in a 
fixed sector with topological charge $Q$, we can fit the results 
for $\la q{'}^{\, 2} \ra$ to the parabola $\chi_{\rm t} V x(1-x)$,
which is predicted by eqs.\ (\ref{p1p2}) and (\ref{q2}).

The simplest case is the topologically trivial sector, $Q=0$,
where $\la q^{2} \ra (x) = \la q{'}^{\, 2}\ra (x)$ is given
by a parabola, which takes its maximum at $x=1/2$, with 
a value of $\la Q^{2} \ra /4$. In the topologically charged 
sectors, $\la q{'}^{\, 2}\ra (x)$ still has the same shape,
whereas $\la q^{2} \ra (x)$ is a parabola that connects 
$\la q^{2} \ra (x=0) =0$ with $\la q^{2} \ra (x=1) = Q^{2}$.
In any sector, the measured data for $\la q^{2} \ra (x)$
can be fitted to the expected parabola. The susceptibility
$\chi_{\rm t}$ is the only fitting parameter,
which is evaluated in this manner.

\section{Results for the quantum rotor}

We first consider a quantum mechanical scalar particle moving 
freely on a unit circle. We deal with periodic boundary 
conditions in Euclidean time. Then there is an integer
winding number for each trajectory, which characterises the
topological sectors of this model. It is also denoted as the 
1d $O(2)$ model, or 1d XY model, and it is related to 
2d $U(1)$ gauge theory \cite{Sinclair}.

Now we assume Euclidean time to be uniformly discretised, and on each 
site $t = 1, 2 \dots L$ there is an angular variable $\phi_{t}$ 
(with $\phi_{t +L} = \phi_{t}$). 
We apply the geometric definition of the topological charge $Q$
\cite{BergLuscher},
\be
Q[\phi ] = \frac{1}{2\pi} \sum_{t=1}^{L} \Delta \phi_{t} \quad , \quad
\Delta \phi_{t} = (\phi_{t+1} -\phi_{t}) ~ {\rm mod} ~ 2 \pi
\in ( -\pi, \pi ] \ , 
\ee
where we define the modulo function such that
$\Delta \phi_{t} $ is the shortest arc length connecting nearest 
neighbour angular variables.

In our numerical studies we simulated three lattice actions:
the standard action, the Manton action \cite{Manton} and 
the constraint action \cite{topact},\footnote{The constraint
action is a special case of a topological lattice action 
\cite{topact,topactU1,topactXY}, which is characterised by the 
property that the action remains invariant under most small 
variations of a configuration.}
\bea
S_{\rm standard}[\phi ] &=& \beta \sum_{t=1}^{L} 
\Big( 1 - \cos (\Delta \phi_{t}) \Big) \nn \\ 
S_{\rm Manton}[\phi ] &=& \frac{\beta}{2} \sum_{t=1}^{L} 
(\Delta \phi_{t})^{2} \nn \\
S_{\rm constraint}[\phi ] &=& \left\{ \begin{array}{cccc}
0 &&& \Delta \phi_{t} < \delta \quad \forall t \\
+ \infty &&& {\rm otherwise.} \end{array} \right.
\eea
The continuum limit is attained at $\beta \to \infty$ or
$\delta \to 0$.

Our simulations were all performed with
a cluster algorithm \cite{Wolff}. Due to its non-local
update steps, the Markov chain changes the topological sector 
frequently, which enables a precise direct measurement of the
topological susceptibility. This result is then confronted with 
the value determined by the slab method, described in Section 2.

We begin with an illustration of measured data for the 
probability product in eq.\ (\ref{p1p2}), as a function of 
the relative slab width $x$ (here the ``slab'' is actually just
an interval in Euclidean time).

\begin{figure}[h!]
\begin{center}
\vspace*{-3mm}
\includegraphics[width=0.45\textwidth,angle=270]{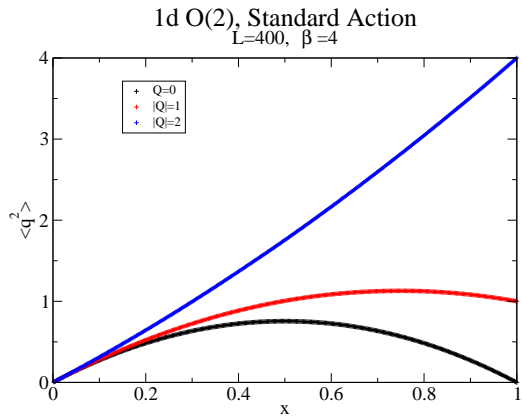}
\vspace*{-2mm}
\caption{The expectation value $\la q^{2} \ra (x)$, measured for the
standard action at $L=400$, $\beta=4$, in the sectors $|Q| =0,\, 1,\, 2$. 
We confirm the predicted parabolic shape behaviour.}
\vspace*{-5mm}
\label{1dO2standard}
\end{center}
\end{figure}

As a first example, Figure \ref{1dO2standard} shows data 
for $\la q^{2} \ra$, as a function of $x$. 
These values were measured with the standard action
at size $L=400$ and $\beta = 4$, which implies a correlation 
length of $\xi \simeq 6.8$. We see that the curves obtained in the 
sectors $Q=0$, $|Q|=1$ and $|Q|=2$ do follow the expected parabolas,
which interpolate $\la q^{2} \ra (x=0) =0$ and
$\la q^{2} \ra (x=1) = Q^{2}$, as predicted in Section 2. 
\vspace*{-3mm}

\begin{figure}[h!]
\begin{center}
\vspace*{-3mm}
\includegraphics[width=0.4\textwidth,angle=270]{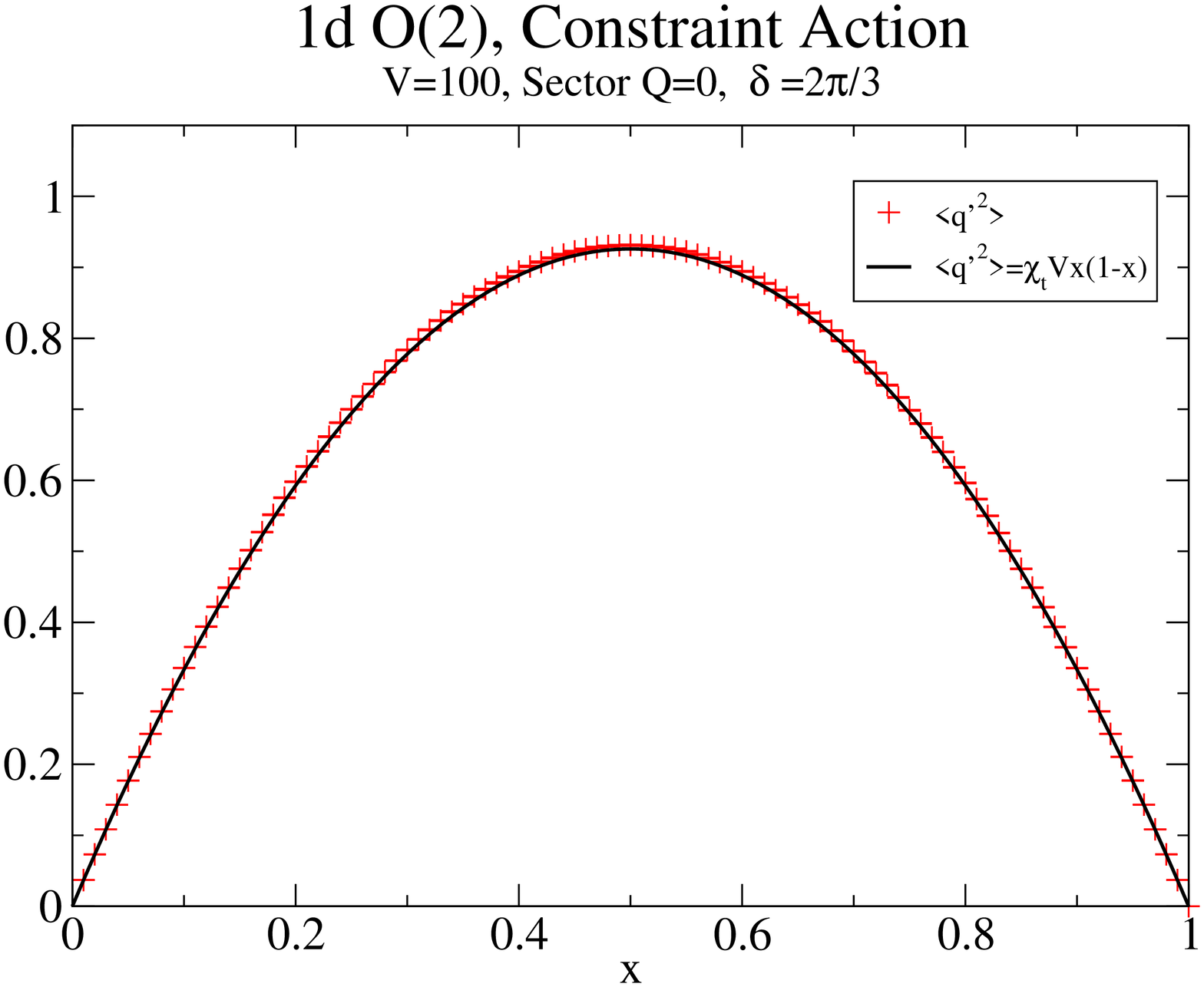}
\vspace*{-3mm} \\
\includegraphics[width=0.4\textwidth,angle=270]{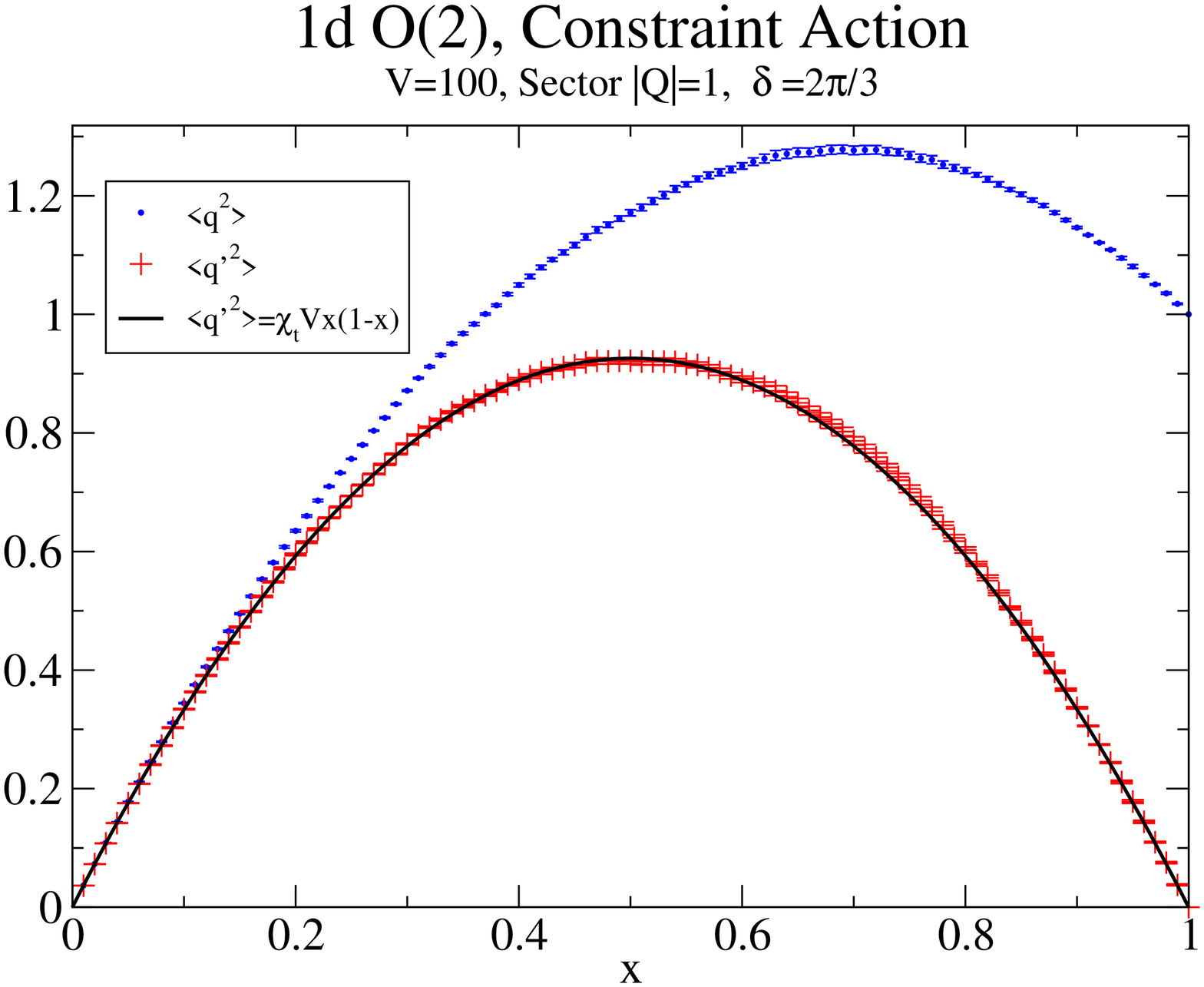} \vspace*{-3mm} \\
\includegraphics[width=0.4\textwidth,angle=270]{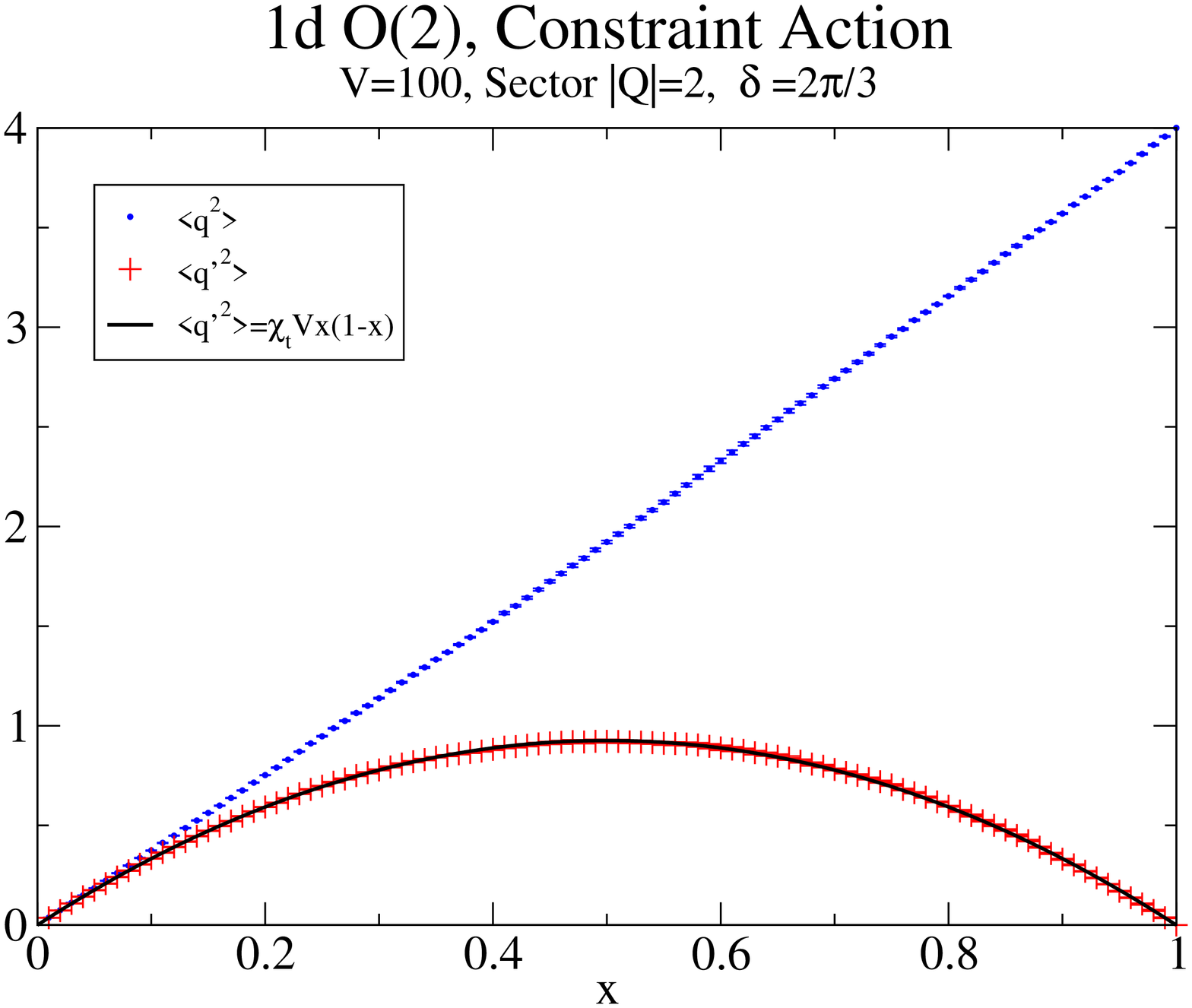} 
\vspace*{-4mm}\\
\caption{The expectation values $\la q^{2} \ra (x)$ and
$\la q{'}^{\, 2} \ra (x)$, measured for the
constraint action at $L=100$, $\delta = 2 \pi /3$, in the sectors 
$|Q| =0,\, 1,\, 2$. For comparison, the black curve shows the
predicted behaviour of $\la q{'}^{\, 2} \ra (x)$, which agrees
very well with the data.}
\label{1dO2constraint}
\vspace*{-5mm}
\end{center}
\end{figure}

Next we give results obtained with the constraint action, 
now at $L=100$ and $\delta = 2 \pi /3$,
where the correlation length is short, $\xi \simeq 1.1$.
Figure \ref{1dO2constraint} shows again the numerical data for 
$\la q^{2} \ra$ as a function of $x$, 
in the sectors $Q=0$, $|Q|=1$ and $|Q|=2$. 
For comparison, we also include the parabola for the
function $\chi_{\rm t} V x(1-x)$, which is predicted for
$\la q{'}^{\, 2} \ra (x)$. Here we insert the directly measured
value of $\chi_{\rm t}$. We see in all cases accurate agreement
between this prediction and the numerical data.\\

\begin{figure}[h!]
\vspace*{-3mm}
\begin{center}
\includegraphics[width=0.45\textwidth,angle=270]{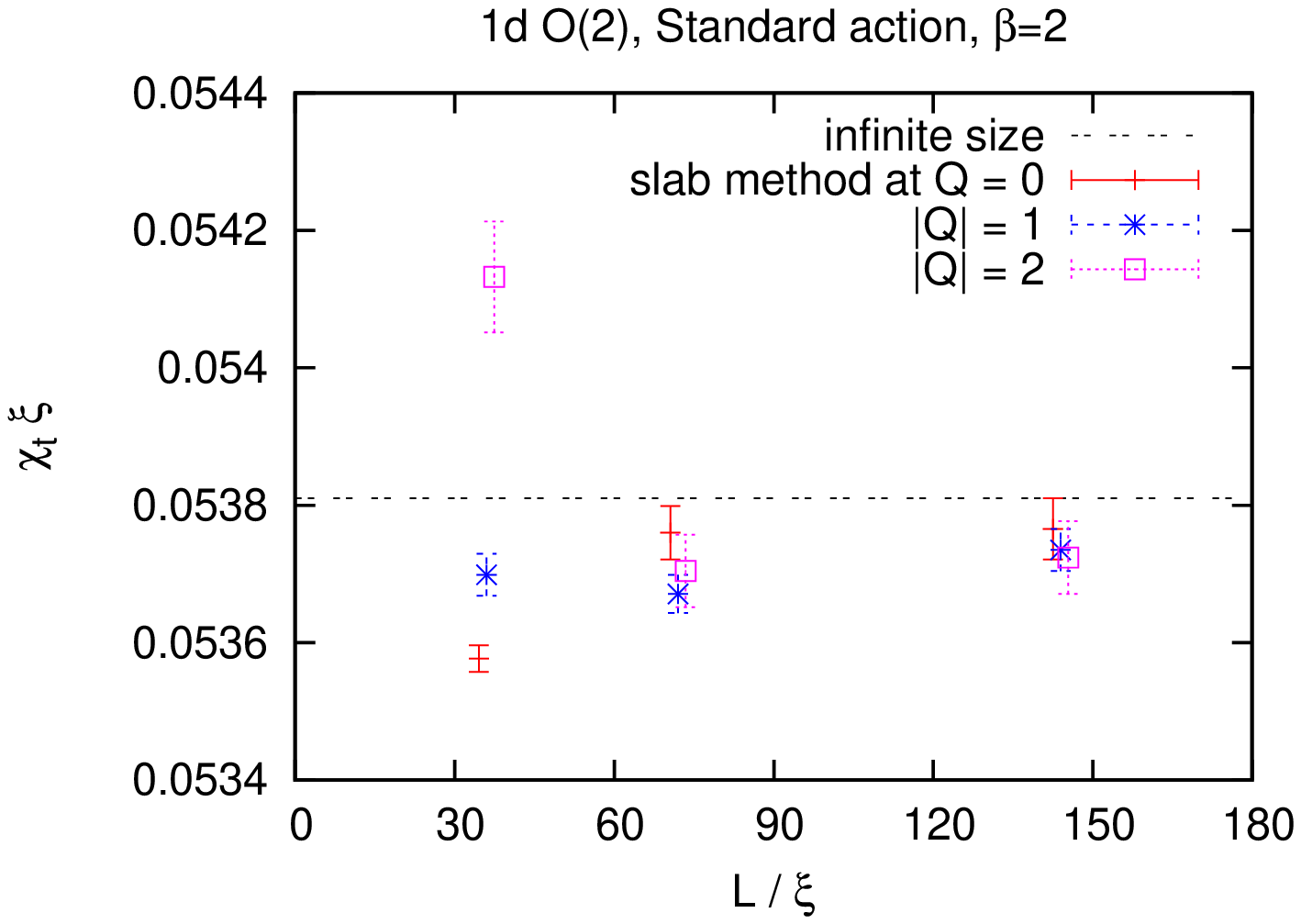}
\includegraphics[width=0.45\textwidth,angle=270]{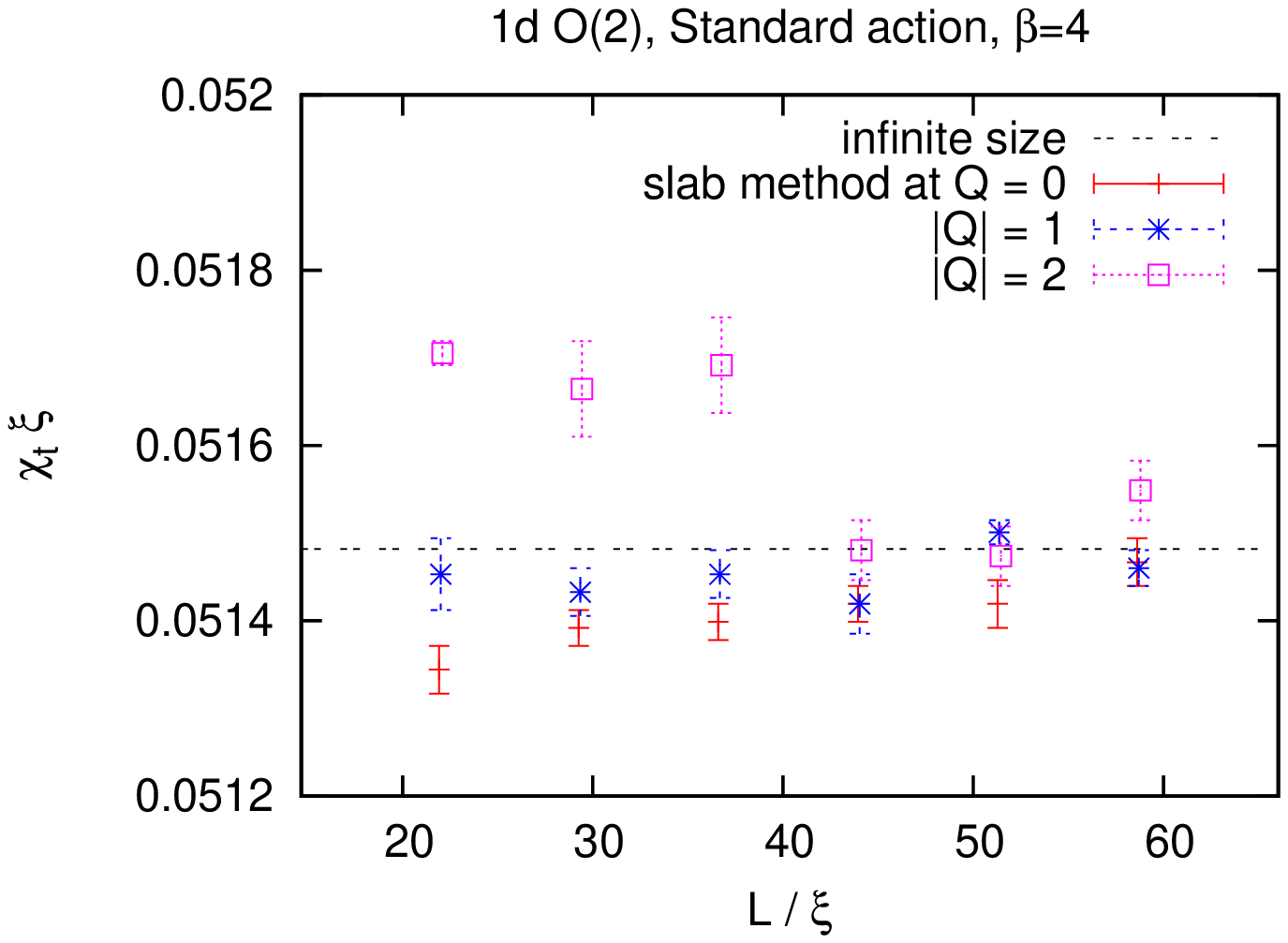} 
\end{center}
\vspace*{-2mm}
\caption{The results for $\chi_{\rm t}$, in units of $1/\xi$,
based on the slab method,
for the standard action at $\beta =2$ (above) and $\beta =4$ (below). 
For an increasing ratio $L/\xi$ they move towards the analytic
result in the thermodynamic limit (dashed line).}
\label{1dO2standardconv}
\end{figure}

\begin{figure}[h!]
\begin{center}
\includegraphics[width=0.4\textwidth,angle=270]{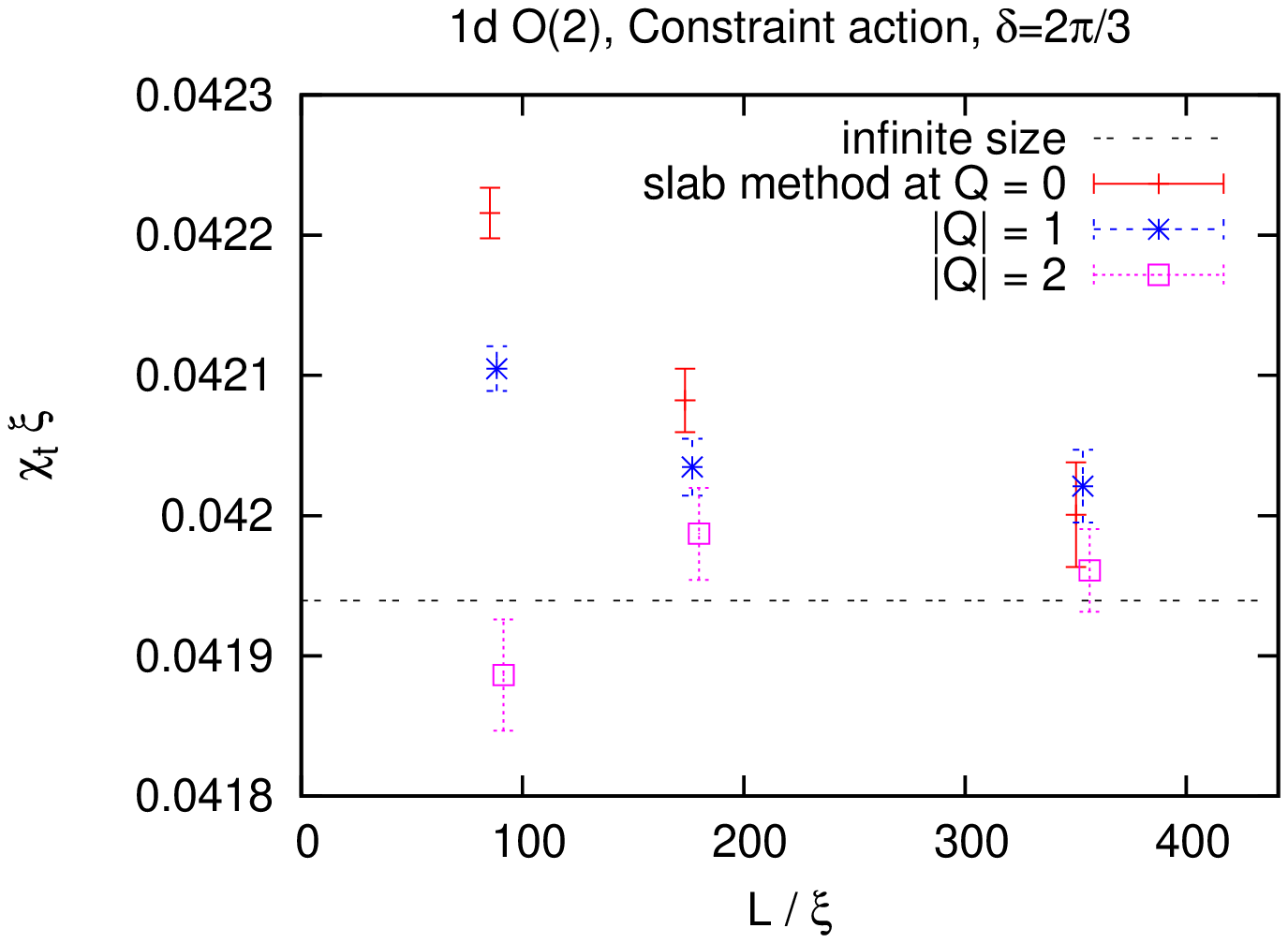}
\includegraphics[width=0.4\textwidth,angle=270]{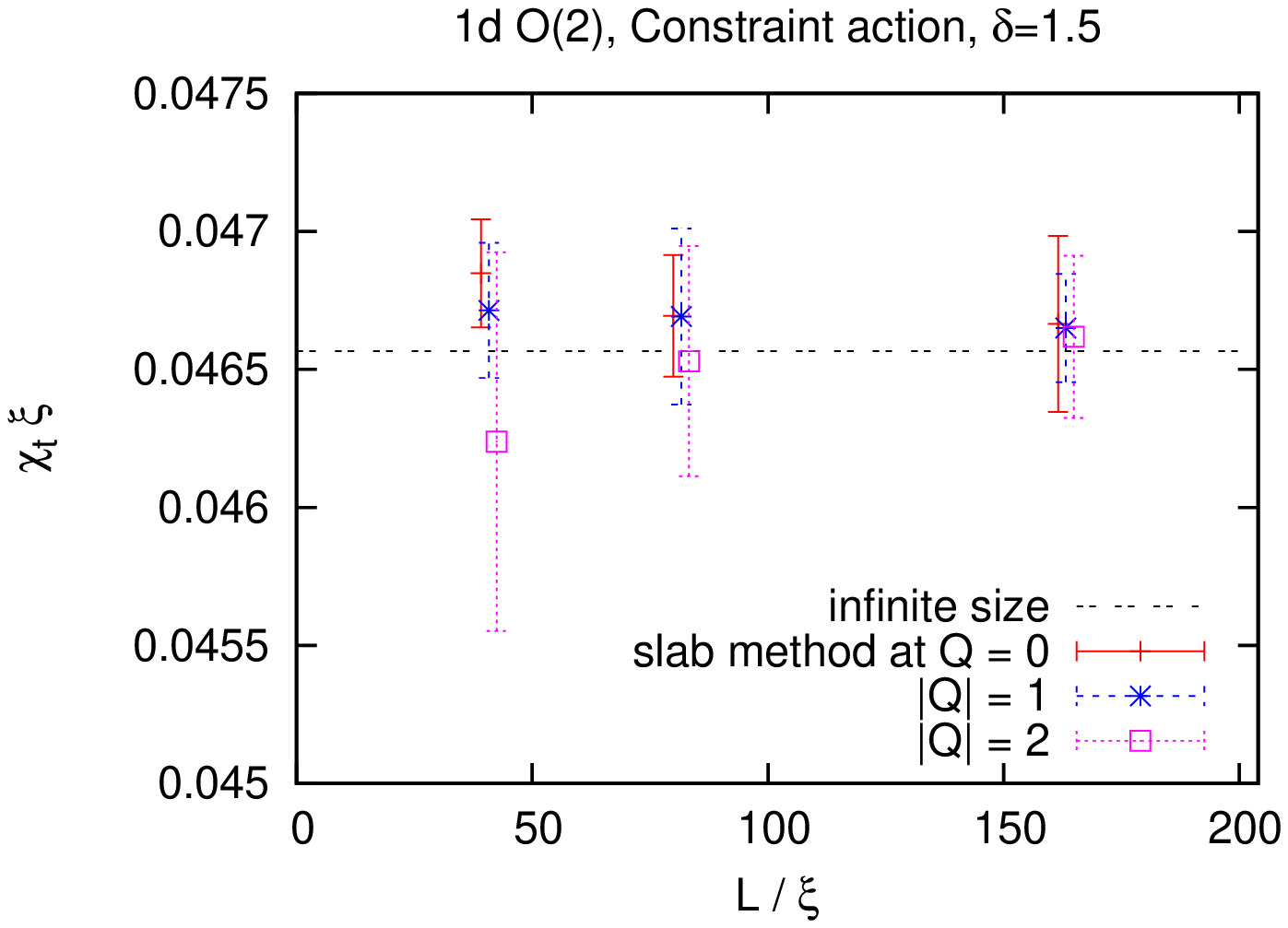}
\includegraphics[width=0.4\textwidth,angle=270]{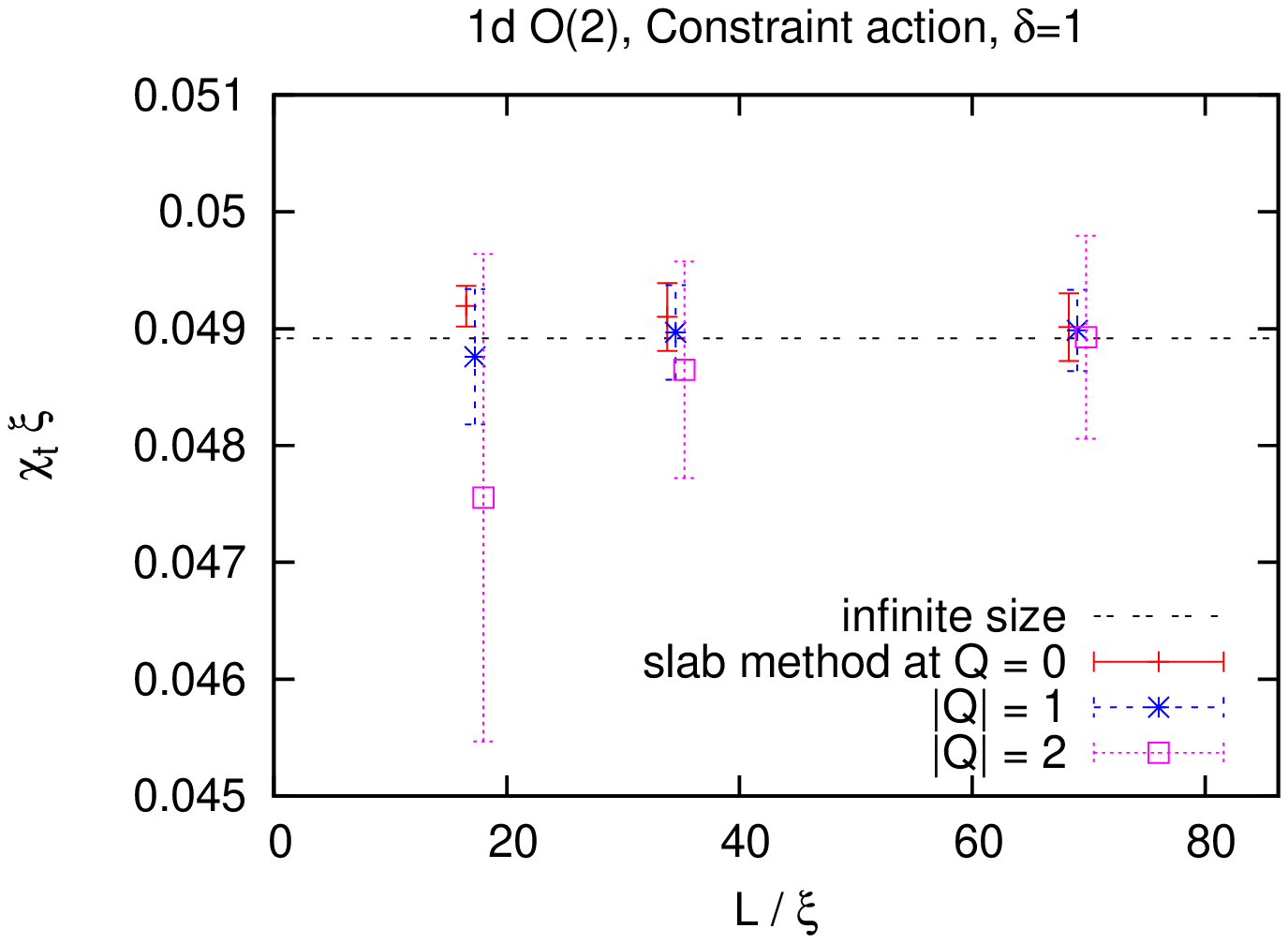}
\end{center}
\caption{Results for $\chi_{\rm t}$, in units of $1/\xi$, 
based on the slab method, for the constraint action at
$\delta =2 \pi /3$, $1.5$ and $1$, as a function of $L/\xi$.}
\label{1dO2constraintconv}
\end{figure}

Let us proceed to a quantitative discussion.
We start again with the standard action, where we consider
$\beta =2$ and $\beta =4$. In a variety of sizes $L$ we
applied the slab method, and evaluated $\chi_{\rm t}$ by a
fit of the data for $\la q{'}^{\, 2} \ra (x)$ to the predicted
parabola. The results are given in Table \ref{1dO2standardtab}.
Figure \ref{1dO2standardconv} shows the corresponding convergence 
of the scaling term $\chi_{\rm t} \xi$, where $\xi$ is the
correlation length.\footnote{Here and in Figures
\ref{1dO2Mantontab} and \ref{1dO2constraintconv} we insert
the analytic values for $\xi$, which are also given in Tables
\ref{1dO2standardtab}, \ref{1dO2Mantontab} and 
\ref{1dO2constrainttab}. They were calculated with the
formulae given in Refs.\ \cite{rot97} and \cite{topact} as functions
of $\beta$ and $\delta$, see also Tables \ref{1dO2standardtab},
\ref{1dO2Mantontab} and \ref{1dO2constrainttab}. 
This refers to $L \to \infty$,
but since $L \gg \xi$ holds in all cases, this is not problematic.}
They are consistently close to the analytical value at infinite 
$L$ (the corresponding formula is given in Ref.\ \cite{rot97}).
That value is in all cases compatible with the directly measured
$\chi_{\rm t}$, which is also included in Table \ref{1dO2standardtab}.
For increasing $L$ the agreement with the slab method results
improves further.\\

\begin{table}[h!]
\begin{center}
\begin{tabular}{|c||c||c|c|c|}
\hline
$L$ & direct & $Q = 0$ & $|Q| = 1$ & $|Q| = 2$ \\
\hline
\hline
\multicolumn{5}{|c|}{$\beta = 2$ \hspace*{1cm}  
$\chi_{{\rm t},L = \infty} = 0.019364$ \hspace*{1cm}  
$\xi = 2.778866$} \\
\hline
100 & 0.019369(6)~ & 0.019280(7)~ & 0.019324(11) & 0.019480(29) \\
\hline
200 & 0.019372(8)~ & 0.019346(14) & 0.019314(10) & 0.019326(19) \\
\hline
400 & 0.019365(10) & 0.019348(16) & 0.019337(11) & 0.019333(19) \\
\hline
\multicolumn{5}{|c|}{$\beta = 4$ \hspace*{1cm}  
$\chi_{{\rm t},L = \infty} = 0.007554$ \hspace*{1cm}
$\xi = 6.814998$} \\
\hline
150 & 0.007554(3) & 0.007534(4) & 0.007550(6) & 0.007587(2) \\
\hline
200 & 0.007557(3) & 0.007541(3) & 0.007547(4) & 0.007581(8) \\
\hline
250 & 0.007549(3) & 0.007542(3) & 0.007550(4) & 0.007585(8) \\
\hline
300 & 0.007560(4) & 0.007545(3) & 0.007545(5) & 0.007554(5) \\
\hline
350 & 0.007554(5) & 0.007545(4) & 0.007557(2) & 0.007553(5) \\
\hline
400 & 0.007549(5) & 0.007552(4) & 0.007551(3) & 0.007564(5) \\
\hline
\end{tabular}
\end{center}
\caption{Explicit results for $\chi_{\rm t}$ by the slab method,
for the standard action at $\beta =2$ and $\beta =4$, as 
illustrated in Figure \ref{1dO2standardconv}.}
\label{1dO2standardtab}
\end{table}

\begin{figure}[h!]
\begin{center}
\includegraphics[width=0.45\textwidth,angle=270]{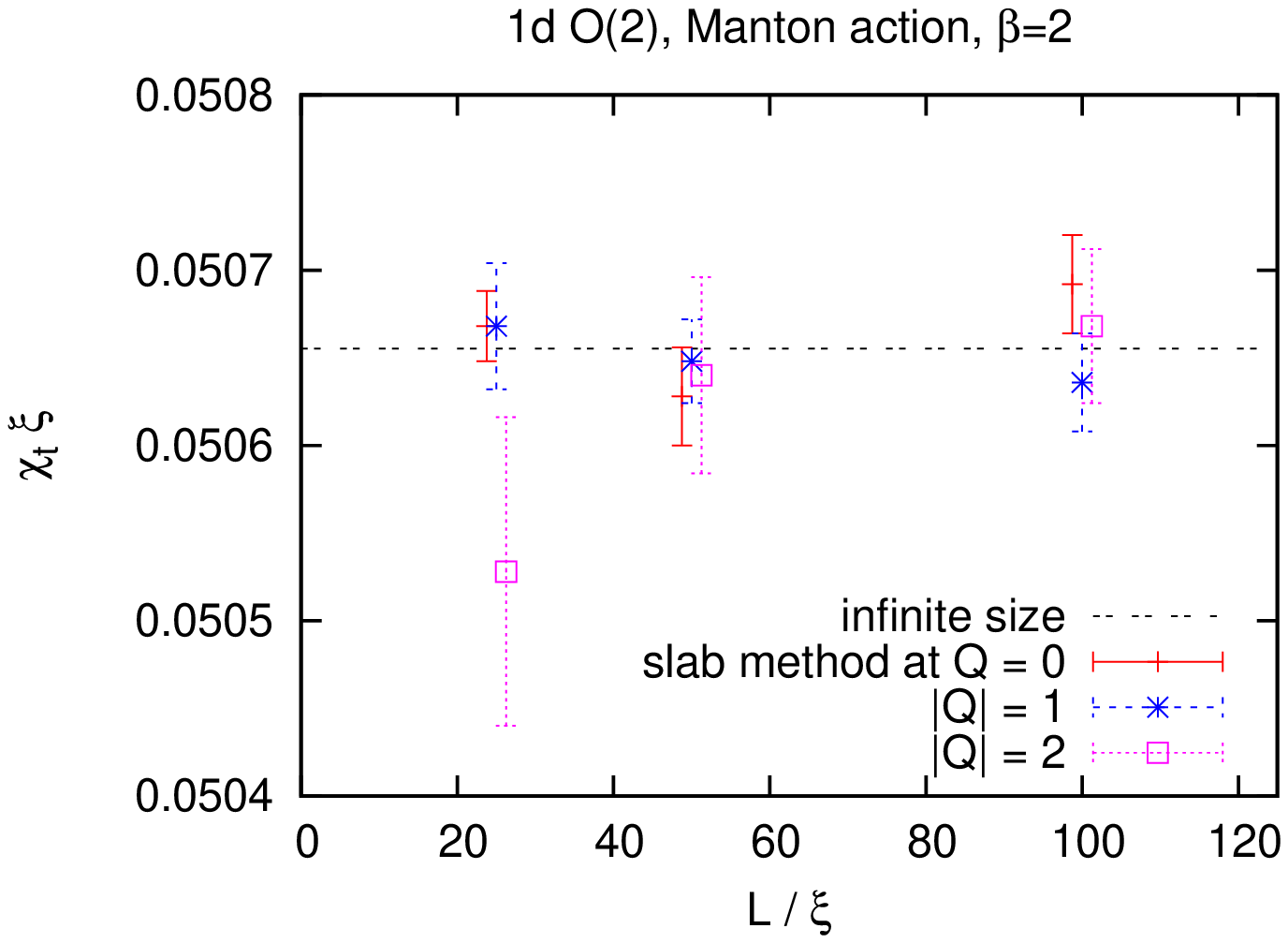}
\includegraphics[width=0.45\textwidth,angle=270]{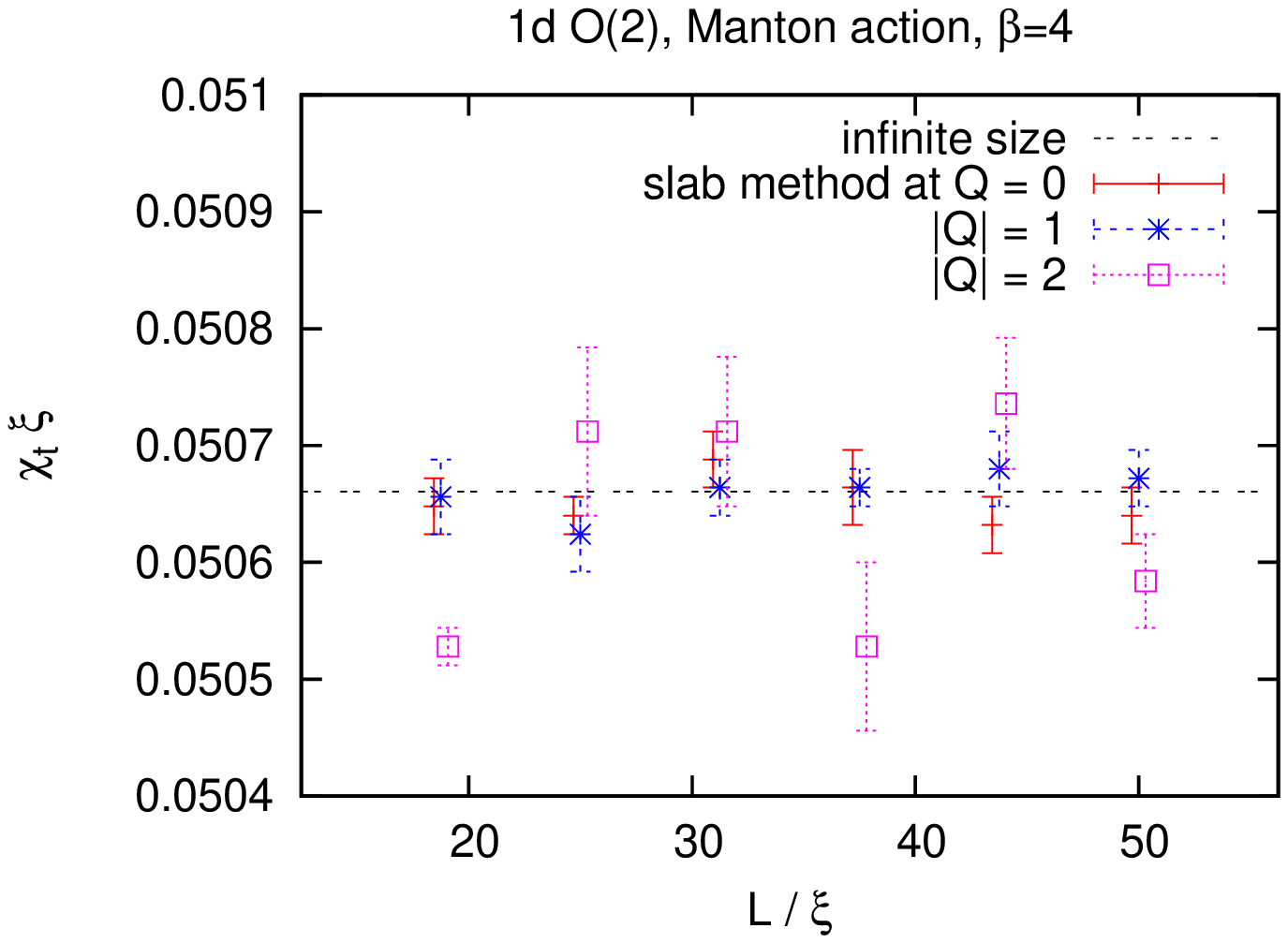} 
\end{center}
\vspace*{-3mm}
\caption{Results for $\chi_{\rm t}$, in units of $1/\xi$,
based on the slab method, for the Manton action at $\beta =2$ 
(above) and $\beta =4$ (below). 
Compared to the standard action, a significantly
smaller ratio $L/\xi$ is sufficient for a very good approximation
to the infinite size result (dashed line).}
\label{1dO2Mantonconv}
\end{figure}

Next we give analogous results for the Manton action, 
Figure \ref{1dO2Mantonconv} and Table \ref{1dO2Mantontab},
and for the constraint action, Figure \ref{1dO2constraintconv}
and Table \ref{1dO2constrainttab}. Qualitatively the same
features are confirmed. Quantitatively we see that the
Manton action --- which is classically perfect \cite{rot97} ---
performs very well regarding the convergence
towards the value at $L = \infty$. For all actions,
the convergence is best for $|Q| \leq 1$, while $|Q|=2$
is affected by somewhat stronger finite size effects.\\

\begin{table}[h!]
\begin{center}
\begin{tabular}{|c||c||c|c|c|}
\hline
$L$ & direct & $Q = 0$ & $|Q| = 1$ & $|Q| = 2$ \\
\hline
\hline
\multicolumn{5}{|c|}{$\beta = 2$ \hspace*{1cm}  
$\chi_{{\rm t},L = \infty} = 0.012663$ \hspace*{1cm} 
$\xi = 4.000321$} \\
\hline
100 & 0.012658(7) & 0.012666(5) & 0.012666(9) & 0.012631(22) \\
\hline
200 & 0.012661(4) & 0.012656(7) & 0.012661(6) & 0.012659(14) \\    
\hline   
400 & 0.012653(4) & 0.012672(7) & 0.012658(7) & 0.012666(11) \\ 
\hline
\multicolumn{5}{|c|}{$\beta = 4$ \hspace*{1cm}  
$\chi_{{\rm t},L = \infty} = 0.006333$ \hspace*{1cm} 
$\xi = 8.000000$} \\
\hline
150 & 0.006330(4) & 0.006331(3) & 0.006332(4) & 0.006316(2)  \\     
\hline
200 & 0.006333(3) & 0.006330(2) & 0.006328(4) & 0.006339(9)  \\
\hline
250 & 0.006335(3) & 0.006336(3) & 0.006333(3) & 0.006339(8)  \\  
\hline
300 & 0.006332(3) & 0.006333(4) & 0.006333(2) & 0.006316(9)  \\      
\hline
350 & 0.006334(4) & 0.006329(3) & 0.006335(4) & 0.006342(7)  \\ 
\hline
400 & 0.006329(3) & 0.006330(3) & 0.006334(3) & 0.006323(5) \\ 
\hline
\end{tabular}
\end{center}
\caption{Explicit results for $\chi_{\rm t}$ by the slab method,
for the Manton action at $\beta =2$ and $\beta =4$, as 
illustrated in Figure \ref{1dO2Mantonconv}.}
\label{1dO2Mantontab}
\end{table}

\begin{table}[h!]
\begin{center}
\begin{tabular}{|c||c||c|c|c|}
\hline
$L$ & direct & $Q = 0$ & $|Q| = 1$ & $|Q| = 2$ \\
\hline
\hline
\multicolumn{5}{|c|}{$\delta = 2 \pi /3$ \hspace*{1cm}  
$\chi_{{\rm t},L = \infty} = 0.037037$ \hspace*{1cm}
$\xi = 1.132367$} \\
\hline
100 & 0.037036(13) & 0.037281(16) & 0.037183(14) & 0.036990(35) \\
\hline
200 & 0.037008(15) & 0.037163(20) & 0.037121(18) & 0.037079(29) \\    
\hline
400 & 0.037042(19) & 0.037091(33) & 0.037109(23) & 0.037056(26) \\ 
\hline 
\multicolumn{5}{|c|}{$\delta = 1.5$ \hspace*{1cm}  
$\chi_{{\rm t},L = \infty} = 0.018998$ \hspace*{1cm}
$\xi = 2.451141$} \\
\hline
100 & 0.018987(8) & 0.019113(8)~ & 0.019058(10) & 0.018864(28) \\
\hline
200 & 0.019011(9) & 0.019050(9)~ & 0.019049(13) & 0.018983(17) \\
\hline
400 & 0.018992(7) & 0.019038(13)& 0.019032(8)~  & 0.019019(12) \\
\hline
\multicolumn{5}{|c|}{$\delta = 1$ \hspace*{1cm}  
$\chi_{{\rm t},L = \infty} = 0.008443$ \hspace*{1cm}
$\xi = 5.793617$} \\
\hline
100 & 0.008443(5) & 0.008491(3) & 0.008416(10) & 0.008208(36) \\
\hline
200 & 0.008445(4) & 0.008475(5) & 0.008452(7)~ & 0.008397(16) \\    
\hline
400 & 0.008439(5) & 0.008460(5) & 0.008455(6)~ & 0.008445(15) \\
\hline
\end{tabular}
\end{center}
\caption{Explicit results for $\chi_{\rm t}$ by the slab method,
for the constraint action at $\delta =2 \pi /3$, $1.5$ and $1$, 
as illustrated in Figure \ref{1dO2constraintconv}.}
\label{1dO2constrainttab}
\end{table}

We also verified that $\la Q \ra$ and the kurtosis term
(which is related to the Binder cumulant),
\be  \label{kurt}
c_{4} = \frac{1}{V} \Big( 3 \la Q^{2} \ra ^{2} - \la Q^{4} \ra \Big) \ ,
\ee
are both compatible with $0$ in all cases (within at most 
$2 \sigma$).\footnote{For these parameters and huge
statistics of $O(10^{9})$ measurements, the error in $c_{4}$
can be reduced to $O(10^{-5})$, and one observes significant
deviations from zero \cite{qqpap}. However, these tiny $c_{4}$ values 
have no influence on the interpretation of the results presented here.} 
The former follows from parity symmetry, and 
a Gaussian $Q$-distribution implies $c_{4} = 0$.
The vanishing of these 
two quantities is an assumption of the slab method. For the kurtosis
this condition is not trivial; it will be considered in detail in
the next section.

Another source of systematic errors are artifacts in $\chi_{\rm t}$
due to the finite size and finite lattice spacing. These artifacts
are discussed in Appendix A.

\section{Results for the 2d Heisenberg model}

We proceed to the 2d $O(3)$ model, where we consider
square lattices with classical spins $\vec e_{x} \in S^{2}$ 
on the sites $x$. In order to define the topological
charge, we cut each plaquette into two triangles (with
an alternating orientation between adjacent plaquettes).
For a triangle with sites $x,\, y,\, z$ we identify the 
(minimal) solid angle spanned by $\vec e_{x}$, $\vec e_{y}$, 
$\vec e_{z}$, including a sign factor for its orientation
(a fully explicit description is given in Ref.\ \cite{topact}).

The sum of the two oriented solid angles within a plaquette, 
divided by $4 \pi$, defines the 
topological charge density $q_{x}$. The total charge of a
spin configuration $[\vec e \, ]$ amounts to
$Q[\vec e \, ] = \sum_{x} q_{x} \in \Z$. It counts how many
times the sum of these solid angles covers the sphere $S^{2}$
with a definite orientation.\\

In analogy to Section 3, we consider three lattice actions,
including an obvious generalisation of the Manton action, which 
was simulated with a new variant of the Wolff cluster algorithm,
\bea
S_{\rm standard}[\vec e \, ] &=& \beta \sum_{x, \mu =1,2} 
(1 - \vec e_{x} \cdot \vec e_{x + \hat \mu} ) \nn \\
S_{\rm Manton}[\vec e \, ] &=& \beta \sum_{x, \mu =1,2} 
\Big( 1 - [{\rm arccos}(\vec e_{x} \cdot \vec e_{x + \hat \mu} ) ]^{2} 
\Big) \nn \\  
S_{\rm constraint}[\vec e \, ] &=& \left\{ \begin{array}{ccc}
0 & & \vec e_{x} \cdot \vec e_{x + \hat \mu} > \cos \delta
\qquad \forall x,\ \mu = 1,2 \\
+ \infty && {\rm otherwise.} \end{array} \right.
\eea

\begin{figure}[h!]
\begin{center}
\includegraphics[width=0.5\textwidth,angle=270]{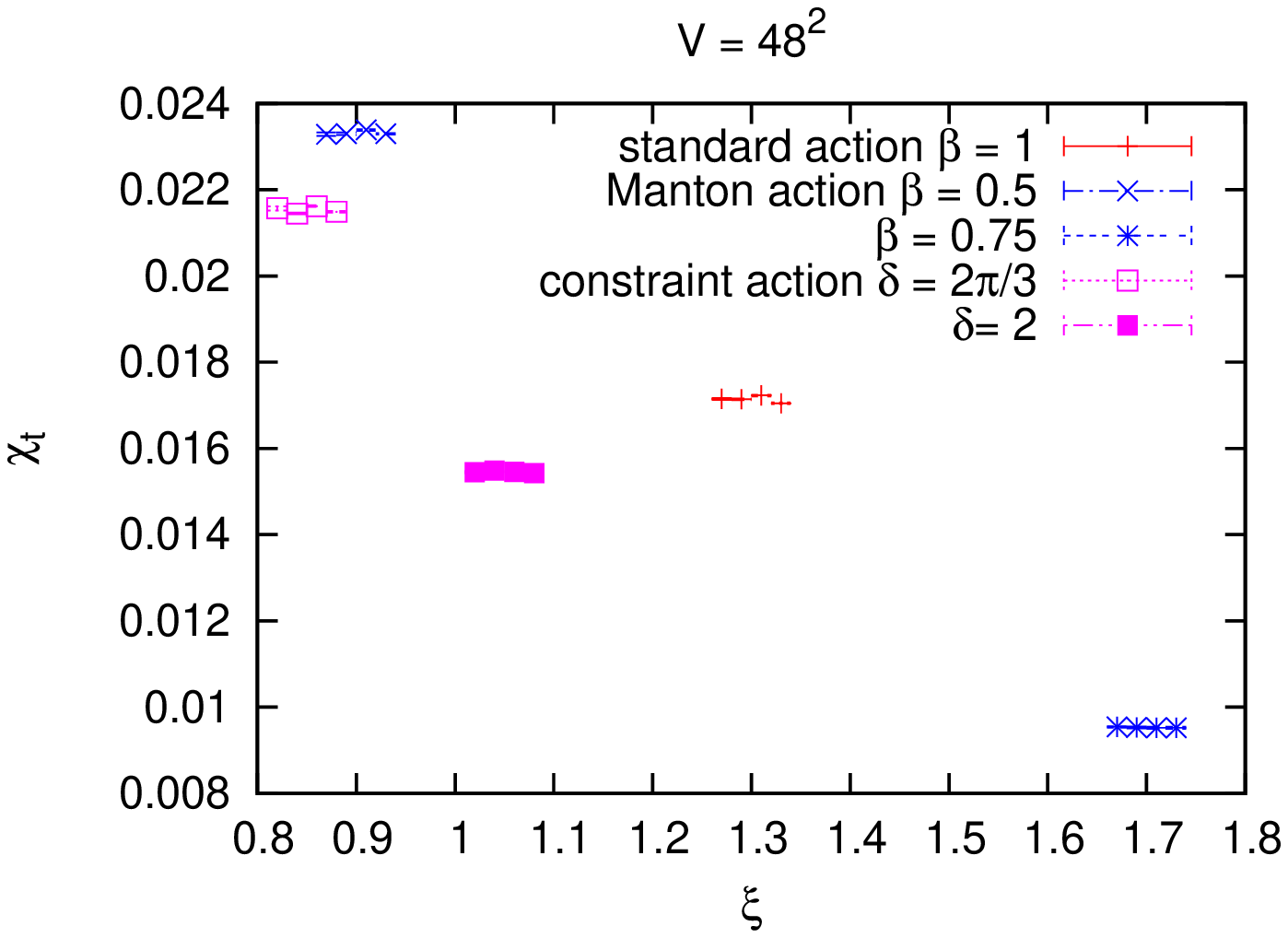}
\includegraphics[width=0.5\textwidth,angle=270]{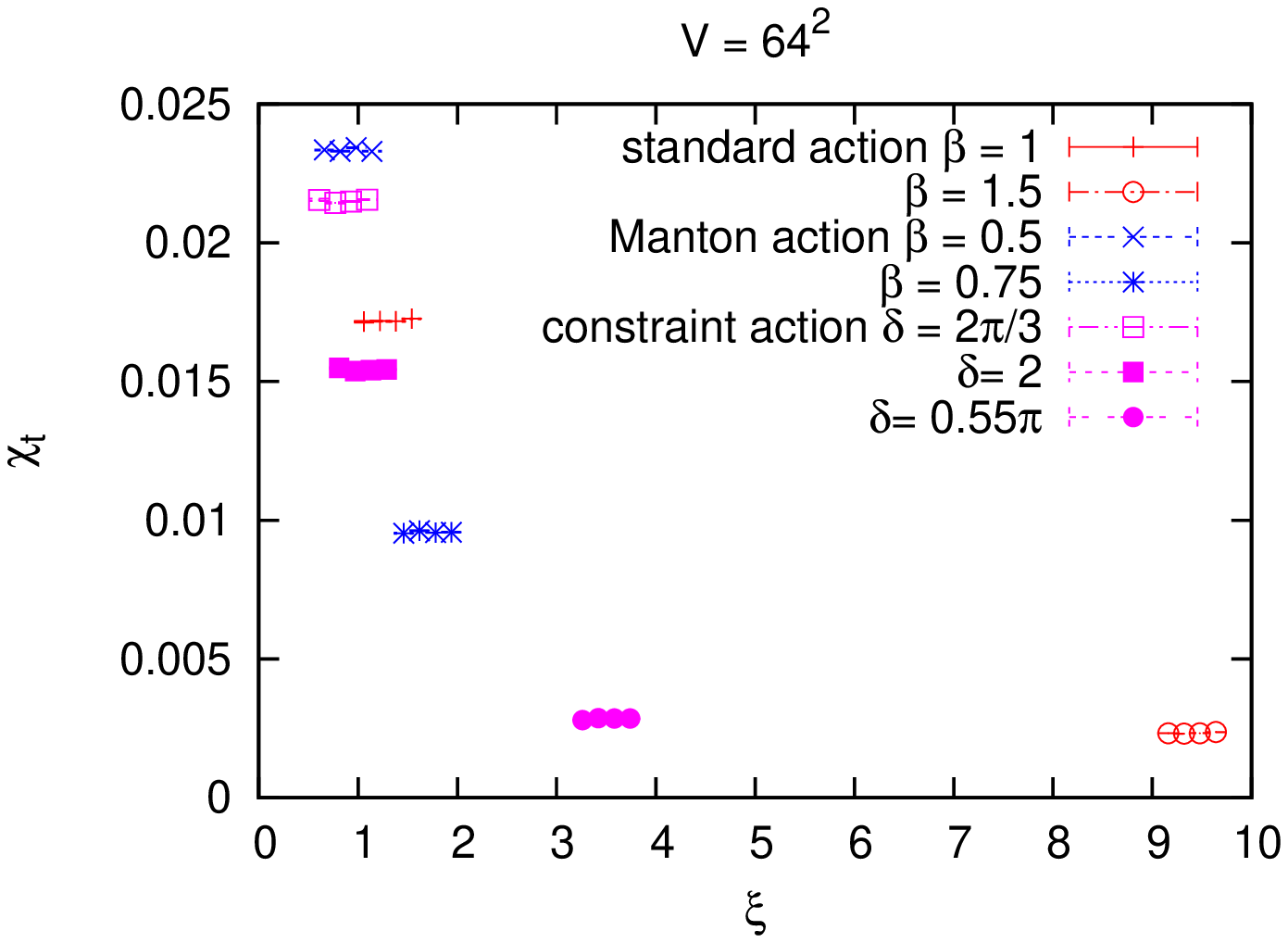}
\caption{Results for the topological susceptibility $\chi_{\rm t}$
in the 2d $O(3)$ model, in the volumes $V=48^{2}$ and
$64^{2}$, for three lattice actions. Each set of results
consists of four data points and shows
(from left to right) the directly measured value, and
the results by the slab method in the sectors $|Q| =0,\, 1,\ 2$.}
\vspace*{-5mm}
\label{2dO3L48L64}
\end{center}
\end{figure}

Regarding the application of the slab method with these lattice 
actions, Figure \ref{2dO3L48L64} shows results obtained in the
lattice volumes $V = 48^{2}$ and $64^{2}$.

We also consider further volumes, including rectangular 
shapes. In fact, it is not obvious if the slab method results
should be compared to $\chi_{\rm t}$ in the entire (periodic)
volume $V$, or to $\chi_{{\rm t},V/2}$ in a (non-periodic) slab of 
size $V/2$. Hence we have measured the latter as well, for 
comparison. The detailed results, along with the corresponding 
correlation length, are given in Tables \ref{2dO3standard},
\ref{2dO3Manton} and \ref{2dO3constraint}.\footnote{In this model,
the apparent scaling quantity $\chi_{\rm t}\, \xi^{2}$ diverges 
logarithmically in the continuum limit (see {\it e.g.}\ Ref.\
\cite{topact}), so here we refer directly to $\chi_{\rm t}$.}
However, for volumes, which are large enough for the slab method 
to work quite well, it turns out that $\chi_{\rm t}$ and
$\chi_{{\rm t},V/2}$ are too close to each other to be distinguished
in light of the slab method results.
\begin{table}[h!]
\hspace*{-1cm}
\begin{tabular}{|c|c||c|c||c||c|c|c|}
\hline
$\beta$ & $V$ & 
$\chi_{\rm t}^{\rm direct}$ & $\chi_{{\rm t}, \, V/2}^{\rm direct}$
& $\chi_{\rm t}^{Q = 0}$ & $\chi_{\rm t}^{|Q| = 1}$ & $\chi_{\rm t}^{|Q| = 2}$ \\
\hline
\multirow{2}{*}{1} & $48^{2}$ 
& 0.01715(2) & 0.01718(2) &
0.01714(2) & 0.01723(2) & 0.01705(2) \\
 & $64^{2}$ 
& 0.01716(3) & 0.01717(2) &
0.01718(2) & 0.01717(2) & 0.01726(2) \\
\hline
\multirow{6}{*}{1.5} & $64^{2}$ 
& 0.002319(3) & 0.002354(2) &
0.002305(5)~ & 0.002322(5) & 0.002361(5) \\
 & $80^{2}$ 
& 0.002332(5) & 0.002356(3) &
0.002327(4)~ & 0.002327(4) & 0.002358(4) \\
 & $64 \times 128$ 
& 0.002332(4) & 0.002344(3) &
0.002304(3)~ & 0.002321(2) & 0.002324(3) \\
 & $96^{2}$ 
& 0.002341(3) & 0.002359(3) &
0.002333(3)~ & 0.002337(3) & 0.002352(3) \\
 & $128^{2}$ 
& 0.002327(3) & 0.002347(3) &
0.002390(10) & 0.002381(9) & 0.002371(9) \\
 & $128 \times 256$ 
& 0.002334(4) & 0.002342(3) &
0.002353(1)~ & 0.002351(1) & 0.002366(1) \\
\hline
\end{tabular}
\caption{Results for the topological susceptibility $\chi_{t}$
in the 2d $O(3)$ model with the standard action. We consider a variety
of volumes at $\beta =1$ and at $\beta =1.5$, with a correlation length
of $\xi = 1.3$ and $9.4$, respectively. The results for $V= 48^{2}$
and $64^{2}$ are illustrated in Figure \ref{2dO3L48L64}.}
\label{2dO3standard}
\end{table}

\begin{table}[h!]
\begin{tabular}{|c|c||c|c||c||c|c|c|}
\hline
$\beta$ & $V$ & 
$\chi_{\rm t}^{\rm direct}$ & $\chi_{{\rm t}, \, V/2}^{\rm direct}$
& $\chi_{\rm t}^{Q = 0}$ & $\chi_{\rm t}^{|Q| = 1}$ & $\chi_{\rm t}^{|Q| = 2}$ \\
\hline
\multirow{2}{*}{0.5} & $48^{2}$
& 0.02329(4) & 0.02333(3) & 0.02330(3) & 0.02339(2) & 0.02330(2) \\
 & $64^{2}$ 
& 0.02335(2) & 0.02335(3) & 0.02330(2) & 0.02343(2) & 0.02330(2) \\
\hline
\multirow{2}{*}{0.75} & $48^{2}$
& 0.00954(2) & 0.00956(1) & 0.00953(2) & 0.00952(2) & 0.00952(2) \\
 & $64^{2}$ 
& 0.00953(2) & 0.00957(1) & 0.00962(2) & 0.00955(1) & 0.00957(2) \\
\hline
\end{tabular}
\caption{Results for the topological susceptibility $\chi_{t}$
in the 2d $O(3)$ model with the Manton action. We consider two
volumes at $\beta = 0.5$ and at $\beta =0.75$, with a correlation length
of $\xi = 0.9$ and $1.7$, respectively. These results are 
illustrated in Figure \ref{2dO3L48L64}.}
\label{2dO3Manton}
\end{table}

\begin{table}[h!]
\hspace*{-1cm}
\begin{tabular}{|c|c||c|c||c||c|c|c|}
\hline
$\delta$ & $V$ & 
$\chi_{\rm t}^{\rm direct}$ & $\chi_{{\rm t}, \, V/2}^{\rm direct}$
& $\chi_{\rm t}^{Q = 0}$ & $\chi_{\rm t}^{|Q| = 1}$ & $\chi_{\rm t}^{|Q| = 2}$ \\
\hline
\multirow{2}{*}{$2\pi /3$} & $48^{2}$
& 0.02157(5) & 0.02155(2) & 0.02145(2) & 0.02162(2) & 0.02149(2) \\
 & $64^{2}$ 
& 0.02144(4) & 0.02151(3) & 0.02144(2) & 0.02149(2) & 0.02156(2) \\
\hline
\multirow{2}{*}{2} & $48^{2}$
& 0.01545(2) & 0.01547(2) & 0.01549(2) & 0.01546(1) & 0.01543(2) \\
 & $64^{2}$ 
& 0.01549(3) & 0.01547(2) & 0.01538(2) & 0.01541(1) & 0.01543(1) \\
\hline
\multirow{6}{*}{$0.55 \, \pi$} 
& $16^{2}$ & 0.002445(6) & 0.002663(4) &
0.002168(65) & 0.003806(14) & 0.005211(27) \\
& $32^{2}$ & 0.002795(4) & 0.002873(3) &
0.002816(17) & 0.002888(16) & 0.003146(15) \\
& $64^{2}$ & 0.002797(4) & 0.002835(4) &
0.002859(6)~ & 0.002853(7)~ & 0.002850(7)~ \\
& $64 \times 128$ & 0.002795(4) & 0.002816(3) &
0.002826(3)~ & 0.002812(3)~ & 0.002810(3)~ \\
& $96^{2}$ & 0.002792(5) & 0.002818(3) &
0.002831(4)~ & 0.002835(4)~ & 0.002843(4)~ \\
& $128^{2}$ & 0.002783(6) & 0.002805(3) &
0.002837(3)~ & 0.002828(3)~ & 0.002819(3)~ \\
\hline
\end{tabular}
\caption{Results for the topological susceptibility $\chi_{t}$
in the 2d $O(3)$ model with the constraint action. We consider a variety
of volumes at $\delta = 2 \pi /3$, $2$ and $0.55 \pi$, with a correlation 
length of $\xi = 0.85$, $1.05$ and $3.5$, respectively. 
The results for $V= 48^{2}$ and $64^{2}$ are illustrated in Figure 
\ref{2dO3L48L64}.}
\label{2dO3constraint}
\end{table}

\begin{figure}[h!]
\begin{center}
\hspace*{-5mm}
\includegraphics[width=0.365\textwidth,angle=270]{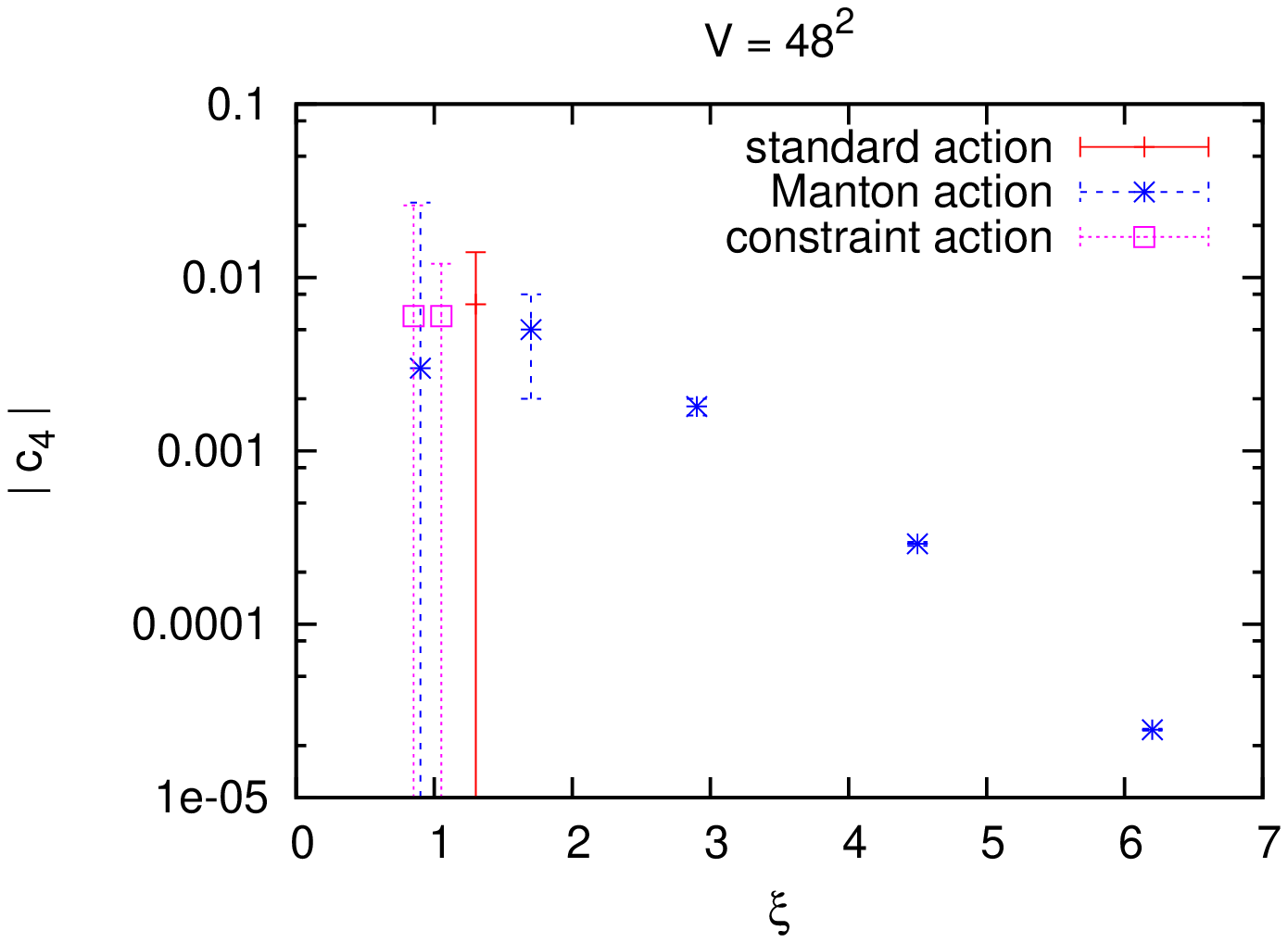}
\hspace*{-5mm}
\includegraphics[width=0.365\textwidth,angle=270]{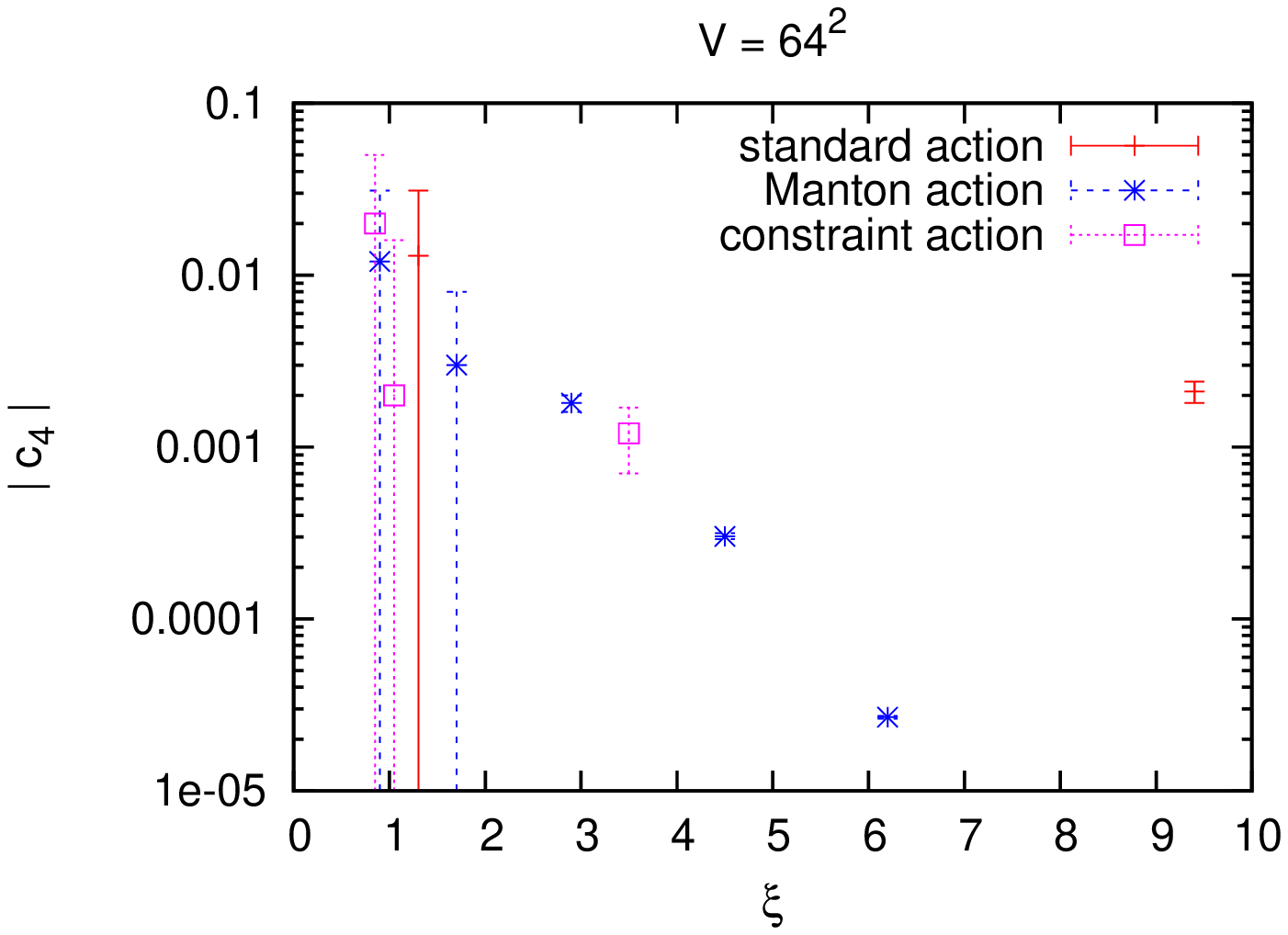}
\caption{The absolute value of the kurtosis, 
given in eq.\ (\ref{kurt}), in the 2d $O(3)$ model. 
For each of the three actions, and at a fixed correlation
length $\xi$, we display the results in $V=48^{2}$ (on the left), and
$V=64^{2}$ (on the right). We see that the volume hardly affects 
the value of $|c_{4}|$, which decreases for increasing correlation 
length $\xi$. This convergence to zero is fastest for the Manton 
action, where we recognise an exponential decrease.}
\vspace*{-5mm}
\label{2dO3kurt}
\end{center}
\end{figure}

The assumption of a Gaussian distribution of the topological charges
is essential for the viability of this method.
Figure \ref{2dO3kurt} shows that the kurtosis (\ref{kurt}) --- as 
a measure for the deviation from a Gaussian --- decreases rapidly as we 
approach the continuum limit; for the Manton action we see an 
exponential decrease of $|c_{4}|$ as $\xi$ increases. 
Thus a non-zero value of $|c_{4}|$ appears as a lattice 
artifact, and not as a finite size effect;
this observation agrees qualitatively with a discussion in Ref.
\cite{qqpap}. This artifact is much less suppressed for the 
standard action and the constraint action.
Regarding the application of the slab method with these lattice 
actions, Figure \ref{2dO3L48L64} shows results obtained in the
lattice volume $V = 48^{2}$ and $64^{2}$.
The systematic errors of the slab method appear as tiny deviations
of $\chi_{\rm t}$ extracted from $|Q|= 0,\, 1$ and $2$ (three
rightmost points in each set of four) from the directly measured
value (leftmost point).

\section{Conclusions}

We have tested an unconventional method for the numerical measurement 
of the topological susceptibility $\chi_{\rm t}$, based on the
division of the volume into slabs. This method is applicable
even if configurations in only one topological sector are available.

Our study shows that --- under suitable conditions --- this method 
works very well. Its statistical accuracy is comparable to the precision 
in the absence of ``topological slowing down'', {\it i.e.}\ the slab 
method is not affected by the freezing of the topological sectors.

In particular, we obtained in the 2d $O(3)$ model
results for $\chi_{\rm t}$, which are correct on the 
percent level, and in the 1d $O(2)$ far beyond. 
This precision was revealed by large statistics of
$O(10^{7})$ measurements, which would not be accessible in higher
dimensional models like QCD; in this sense, the accuracy of
the slab results are fully satisfactory.

The slab method is most successful at small topological 
charges, $|Q| \leq 1$, whereas its application
in the sector $|Q|=2$ is more tedious, since the 
finite size effects are worse.
However, even at $|Q| \leq 1$ the finite size effects are
highly persistent. In fact, other methods and formulae
show that these effects are only suppressed by a power
series in $1/V$ for topologically fixed measurements 
\cite{othermethods,qqpap}. 

As an illustrative example of possible systematic effects, 
we refer to the results at $\delta = 0.55\pi$ in 
Table \ref{2dO3constraint}: even the smallest volume, $V=16^{2}$, 
corresponds to the ratio $L/\xi \simeq 4.6$, where one would expect 
usual  ({\it i.e.}\ exponentially suppressed) finite size effects to
be mostly eliminated. The results for the slab method, 
however, are strongly distorted --- they improve significantly
at $L=32$ (and for $|Q|=2$ it even takes $L=64$).

This illustrates the general feature: at fixed $\beta$ or $\delta$,
{\it i.e.}\ at approximately constant $\xi$, the slab results
converge for increasing volume. In order to assure that they
do converge to the (vicinity of the) correct value, we further
have to require the topological charges $Q$ to be
(approximatively) Gauss-distributed (that also ensures a good fit
to formula \ref{p1p2}). Therefore we also studied the
kurtosis, which may deviate from zero (indicating a deviation
from a Gaussian) as a lattice artifact. It is suppressed when
$\xi$ increases (in lattice units), with a rate that depends on 
the lattice action; for the Manton action we observed a 
particularly fast, exponential suppression.

Hence $\xi$ should be sufficiently large to control that requirement,
and the volume should be large enough to obtain $L/\xi = O(10)$; then
the slab method can be expected to provide the correct $\chi_{\rm t}$
value on the percent level.

\ \\
\noindent
{\bf Acknowledgements} \\ We thank Irais Bautista and Christoph P.\
Hofmann for inspiring discussions. This work was supported in part 
by the Mexican {\it Consejo Nacional de Ciencia y Tecnolog\'{\i}a} 
(CONACYT) through projects CB-2010/155905 and 
CB-2013/222812, and by DGAPA-UNAM, grant IN107915.

\appendix

\section{Artifacts in the topological susceptibility}

An essential point for the quality of the results obtained by the
slab method is the precision of $\chi_{\rm t}$ as measured
in the sub-volumes. In this appendix we discuss the artifacts
that occur: Sub-Appendix A.1 deals with the finite size effects, 
which depend on the ratio $\xi/L$, and which even occur in the
continuum formulation. Sub-Appendix A.2 considers the lattice artifacts,
as a function of $a/L$ (where $a$ is the lattice spacing, and $L$ is
now the dimensional size). Both considerations are performed
in the analytically tractable case of the quantum rotor;
Sub-Appendix A.2 captures the standard and the Manton action.
For simplicity we deal with periodic boundary conditions throughout,
although they are partly open in the sub-volumes, so we assume implicitly 
that $L$ is large enough for this distinction to be negligible.

\subsection{Finite size effects}

While the slab method is plagued by particularly persistent
finite size effects, the true $\chi_{\rm t}$ value tends to converge
exponentially as the volume is enlarged. Here we discuss this 
convergence for the case of the 1d $O(2)$ model.

\begin{figure}[h!]
\center
\includegraphics[angle=270,width=0.65\linewidth]{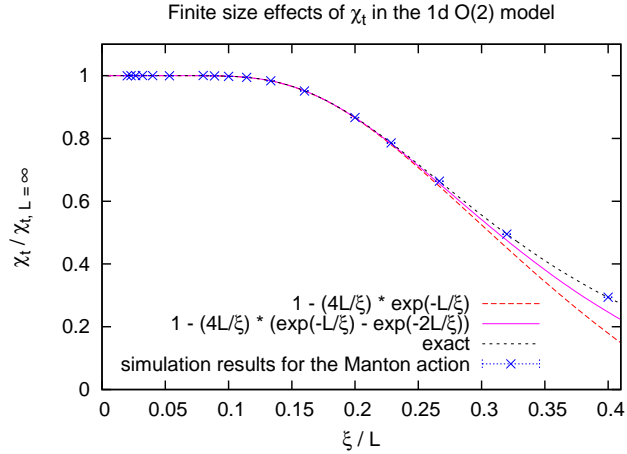} 
\caption{The ratio $\chi_{\rm t}/ \chi_{{\rm t},\, L = \infty}$ as a function
of $\xi /L$. We show the leading and next-to-leading 
finite size correction in approximation (\ref{finale}), the exact
result and simulations results for the Manton action at $\beta =4$.}
\label{correct}
\end{figure}

In the continuum formulation, with the action
$S [\varphi ] = \frac{\beta}{2} \int_{0}^{L} \dot \phi^{2} \ dt$,
the topological susceptibility takes the form \cite{rot97}
\be
\chi_{\rm t} = \frac{1}{L} \la Q^{2} \ra =  \frac{1}{L}
\frac{\sum_{Q \in \Z} Q^{2} e^{- \alpha \pi Q^{2} }}
{\sum_{Q \in \Z} e^{-\alpha \pi Q^{2}}} \ , \qquad
\alpha := 2 \pi \beta /L \ .
\label{fullsum}
\ee
It can be written in terms of a Jacobi $\theta$-function,
\be \label{chitJ}
\chi_{\rm t} = - \frac{1}{\pi L} \frac{d}{d \alpha} \ln \theta (\alpha ) 
\ , \qquad \theta (\alpha ) := \sum_{Q \in \Z} e^{- \alpha \pi Q^{2}} \ .
\ee
By applying the Jacobi identity,
\be  \label{Jac}
\theta (\alpha ) 
= \frac{1}{2 \pi \sqrt{\alpha}} \ \theta (1/\alpha ) \ ,
\ee
we obtain
\be  \label{Japprox}
\theta (\alpha )\vert_{0 < \alpha \ll 1} \simeq 
\frac{1}{2 \pi \sqrt{\alpha}} \ \Big( 1 + 2 e^{-\pi/\alpha}
+ 2 e^{- 4 \pi/\alpha} \dots \Big) \ .
\ee
Inserting approximation (\ref{Japprox}) into formula (\ref{chitJ}),
we arrive at 
\be
\chi_{\rm t} \simeq \frac{1}{(2 \pi)^{2} \beta} + \frac{L}{2 \pi^{2} \beta^{2}} 
\Big( - e^{-L/(2 \beta)} + e^{-L/ \beta} \Big) \ .
\ee
With the correlation length $\xi = 2 \beta$ \cite{rot97},
this corresponds to the ratio
\be  \label{finale}
\frac{\chi_{\rm t}}{\chi_{{\rm t},\, L = \infty}} \simeq 
1 + \frac{4L}{\xi} \ \Big( - e^{-L/\xi} + e^{- 2 L/\xi} \Big) \ .
\ee
Thus we see explicitly the exponentially suppressed finite size
corrections. Figure \ref{correct} illustrates this formula,
and compares it to the exact result (numerical summation of the
series in eq.\ (\ref{fullsum})), which agrees with simulation data for 
the Manton action at $\beta =4$ (where lattice artifacts are
practically erased).

\subsection{Lattice artifacts}

We begin with general properties of a field theory in a periodic
Euclidean volume $V$, where the configurations of the field $\phi$
are divided into topological sectors. In presence of a vacuum angle
$\theta$ (and with $\hbar =1$), the partition function 
and the topological susceptibility can be written as\footnote{Occasionally
we omit the argument $\theta$ in the partition function and (below) in
the energy eigenvalues.}
\be
Z (\theta ) = \int D \phi \ e^{-S[\phi ] - \ri Q[\phi ] \theta} \ ,
\qquad 
\chi_{\rm t} =  
\frac{1}{V} \Big[ - \frac{Z''}{Z} + \Big( \frac{Z'}{Z} \Big)^{2} \Big] \ .
\label{chit1}
\ee

In the Hamilton formulation, the partition function reads
$$
Z(\theta ) = {\rm Tr} \, e^{-\beta \hat H (\theta)} =
\sum_{n} e^{-\beta E_{n}(\theta )} \ , 
$$
where $\beta$ is the inverse temperature, and we assume a discrete
energy spectrum. Inserting eq.\ (\ref{chit1}) leads to
\be
\chi_{\rm t} = \frac{1}{Z} \frac{\beta}{V} \sum_{n} 
[ E_{n}'' - \beta (E_{n}')^{2} ] \, e^{- \beta E_{n}}
+ \frac{1}{Z^{2}} \frac{\beta^{2}}{V} 
\Big( \sum_{n} \, E_{n}' \, e^{- \beta E_{n}} \Big)^{2} \ .
\ee

Now we focus on $d=1$, and replace $V$ and $\beta$ both by $L$
(the periodicity range in Euclidean time). More specifically, we 
consider again the quantum rotor, where $L$ also represents the moment 
of inertia. We deal with a lattice of spacing $a$, and periodicity 
over $N$ sites, $L = N a$.
A consideration of the transfer matrix yields \cite{rot97}
$$
e^{-a E_{n}} = \sqrt{\frac{L}{2 \pi a}} \int_{-\pi}^{\pi} d \vp \
\exp \Big( - \frac{L}{a} f(\vp ) - \ri \Big( n - \frac{\theta}{2 \pi}
\Big) \vp \Big) \doteq 
\sqrt{\frac{L}{2 \pi a}} \, I_{n}(\theta ) \ ,
$$
where $n \in \Z$, and
$$
f (\vp ) = \left\{ \begin{array}{cccc}
1 - \cos \vp &&& {\rm standard~action} \\
\frac{1}{2} \vp^{2} &&& {\rm Manton~action.} \end{array} \right.
$$
In either case, $f(\vp )$ is even, such that $E_{n}'(0) = 0$,
in accordance with $\la Q \ra_{\theta =0} = 0$, and some algebra leads to
\bea
\chi_{{\rm t}, \theta = 0}
&=& \left. \frac{1}{a} \frac{
\sum_{n} [- I_{n}'' I_{n} + (N+1) (I_{n}')^{2} ] I_{n}^{N-2} }
{\sum_{m} I_{m}^{N} } \right|_{\theta = 0} \ , \nn \\
{\rm with} && 
I_{n}(0) = \int_{-\pi}^{\pi} d \vp \ e^{-f(\vp )L/a} \cos (n \vp ) \nn \\
&& I_{n}'(0) = \frac{1}{2\pi} \int_{-\pi}^{\pi} d \vp \ e^{-f(\vp )L/a} 
\vp \sin (n \vp ) \nn \\
&& I_{n}''(0) = - \frac{1}{(2\pi )^{2}}
\int_{-\pi}^{\pi} d \vp \ e^{-f(\vp )L/a} \vp^{2} \cos (n \vp ) \ .
\eea

To proceed, we now have to specify $f(\vp )$. We start with
the {\em Manton action,} and substitute 
$\bar \vp \doteq 
\vp \sqrt{N} $,
\bea
I_{n}(0) &=& \frac{1}{\sqrt{N}} \int_{-\pi \sqrt{N}}^{\pi \sqrt{N}} d \bar \vp \
e^{- \bar \vp^{2}/2} \cos (n \bar \vp / \sqrt{N}) 
\doteq \frac{1}{\sqrt{N}} J_{n}^{(0)} \nn \\
I_{n}'(0) &=& \frac{1}{2 \pi N} \int_{-\pi \sqrt{N}}^{\pi \sqrt{N}} d \bar \vp \
e^{- \bar \vp^{2}/2} \bar \vp \sin (n \bar \vp / \sqrt{N} ) \doteq 
\frac{1}{2 \pi N} J_{n}^{(1)} \nn \\
I_{n}''(0) &=& - \frac{1}{(2 \pi )^{2} N^{3/2}} 
\int_{-\pi \sqrt{N}}^{\pi \sqrt{N}} d \bar \vp \
e^{- \bar \vp^{2}/2} \bar \vp^{2} \cos (n \bar \vp / \sqrt{N} ) 
\doteq \frac{1}{(2 \pi )^{2} N^{3/2}} J_{n}^{(2)} \nn \\
\chi_{{\rm t}, \theta = 0}^{\rm Manton} &=& \frac{1}{(2 \pi )^{2}LN}
\frac{ \sum_{n} [ - J_{n}^{(2)} J_{n}^{(0)} + (N+1) (J_{n}^{(1)})^{2}]
\ (J_{n}^{(0)})^{N-2}} {\sum_{m} (J_{m}^{(0)})^{N}} \ . \quad
\label{Manton}
\eea
If we write $\cos (n \bar \vp / \sqrt{N})$ and
$\sin (n \bar \vp / \sqrt{N} )$ in exponential form, 
and complete the squares in the exponents,
\be  \label{squarexp}
- \frac{1}{2} \bar \vp^{2} + \ri n \vp / \sqrt{N} =
- \frac{1}{2} (\bar \vp - \ri n / \sqrt{N})^{2} - \frac{n^{2}}{2N} \ ,
\ee
we see several types of lattice artifacts, such as the
extra term $ 
-a n^{2} /(2L)$ in the exponent, and the 
incomplete Gauss integrals; they are {\em exponentially} 
suppressed in $a/L$. The substitution of the integration variable 
$
\bar \vp \to \bar \vp - \ri n / \sqrt{N} 
$
corresponds to a shift of the integration contour;
for the incomplete Gauss integral this
is another artifact, which is exponentially suppressed.
In summary, there are no power lattice artifacts for the Manton action.\\

For the {\em standard action} the corresponding formulae read
\bea
\chi_{{\rm t}, \theta = 0}^{\rm standard} &=& \frac{1}{(2 \pi )^{2}a} \frac{
\sum_{n} [\tilde I_{n}^{(2)} \tilde I_{n}^{(0)} + (N+1) (\tilde I_{n}^{(1)})^{2} ] 
(\tilde I_{n}^{(0)})^{N-2} } {\sum_{m} (\tilde I_{m}^{(0)})^{N} } \ , \nn \\
{\rm with} && \tilde I_{n}^{(0)} \doteq  \int_{-\pi}^{\pi} d \vp \
e^{ (\cos \vp -1) N} \cos (n \vp ) \nn \\
&& \tilde I_{n}^{(1)} = \int_{-\pi}^{\pi} d \vp \
e^{ (\cos \vp -1) N} \vp \sin (n \vp ) \nn \\
&& \tilde I_{n}^{(2)} = \int_{-\pi}^{\pi} d \vp \, e^{ (\cos \vp -1) N} \vp^{2}
\cos (n \vp ) \ .
\eea

Regarding the search for {\em power lattice artifacts,} we start
from the Manton action and add the two leading modifications
of the standard action,
$ f(\vp ) \simeq \vp^{2}/2 - \vp^{4}/4! + \vp^{6} /6! \, $.
This modifies the expressions for $I_{n}(0)$, $I_{n}'(0)$ and $I_{n}''(0)$ 
in eq.\ (\ref{Manton}) by a factor 
$$
\Big( 1 + \frac{1}{4! N} \bar \vp^{4} +
\frac{1}{2 (4! N)^{2}} \bar \vp^{8} - \frac{1}{6! N^{2}} \bar \vp^{6} \Big)
$$
under the integrals, 
and after some calculus we obtain
\be
\chi_{{\rm t}, \theta = 0}^{\rm standard} \simeq \chi_{{\rm t}, \theta = 0}^{\rm Manton}
\Big( 1 + \frac{a}{2L} + \frac{13}{24} \frac{a^{2}}{L^{2}} \Big) \ .
\ee
Hence for the standard action $\chi_{{\rm t}, \theta = 0}$
is affected by corrections linear in $a/L$.

This may appear surprising for a bosonic theory, but one has to
keep in mind that $\chi_{{\rm t}, \theta = 0}$ is not a scaling quantity.
The actual scaling artifacts refer to the product $\chi_{\rm t} \xi$.
In the thermodynamic limit they are expressed in powers of
$a/\xi$ \cite{topact},
\be
\chi_{\rm t} \xi = \left\{ \begin{array}{ccc}
\frac{1}{2 \pi^{2}} \Big( 1 + \frac{1}{3} \frac{a^{2}}{\xi^{2}} \dots \Big)
&& {\rm standard~action} \\
\frac{1}{2 \pi^{2}} \Big(
1 - \sqrt{\pi \xi}{a} (1 - 4 /\pi^{2}) e^{- \pi^{2} \xi /(4 a)} + \dots \Big)
&& {\rm Manton~action} \\
\frac{1}{2 \pi^{2}} \Big(
1 - \frac{1}{5} \frac{a}{\xi} + \dots \Big) && {\rm constraint~action.}
\end{array}
\right. \nn
\ee
For the standard action, the linear artifacts cancel in this scaling 
quantity, 
while the Manton action is classically perfect; its scaling artifacts are
exponentially suppressed. As an exotic feature, the constraint action
does have linear artifacts, due to the absence of a derivative term
in the action.\\

We summarise that the finite size effects are exponentially suppressed
in $L/\xi$, whereas the type of lattice artifacts depend on the 
lattice action.

\end{document}